\documentclass[a4paper,UKenglish,cleveref, autoref, thm-restate,nameinlink]{lipics-v2021}

\pdfoutput=1 
\hideLIPIcs  

\title{Elimination Distance to Dominated Clusters} 

\author{Nicole Schirrmacher}{University of Bremen, Germany}{schirrmacher@uni-bremen.de}{https://orcid.org/0000-0002-1740-7478}{}

\author{Sebastian Siebertz}{University of Bremen, Germany}{siebertz@uni-bremen.de}{https://orcid.org/0000-0002-6347-1198}{}

\author{Alexandre Vigny}{Université Clermont Auvergne, Clermont Auvergne INP, LIMOS, CNRS, Clermont-Ferrand, France}{alexandre.vigny@uca.fr}{https://orcid.org/0000-0002-4298-8876}{} 

\authorrunning{N. Schirrmacher and S. Siebertz and A. Vigny} 

\Copyright{Nicole Schirrmacher and Sebastian Siebertz and Alexandre Vigny} 



\category{} 

\relatedversion{} 

\funding{
   This paper is a part of the ANR-DFG project \emph{Unifying Theories for Multivariate Algorithms} (\rm{UTMA}), which has received funding from the German Research Foundation (DFG) with grant agreement No 446200270, and benefited from state aid managed by the ANR under France 2030 referenced ANR-23-IACL-0006.
}

\nolinenumbers 

\EventEditors{John Q. Open and Joan R. Access}
\EventNoEds{2}
\EventLongTitle{50th International Symposium on
Mathematical Foundations of Computer Science (MFCS 2025)}
\EventShortTitle{MFCS 2025}
\EventAcronym{MFCS}
\EventYear{2025}
\EventDate{August 25 – 29, 2025}
\EventLocation{Warsaw, Poland}
\EventLogo{}
\SeriesVolume{42}
\ArticleNo{XX}

\usepackage{algorithm}
\usepackage{algpseudocode}
\usepackage{tikz-cd}
\usetikzlibrary{decorations.pathmorphing}
\usepackage{fontaxes}
\usepackage{trimspaces}
\usepackage{nccfoots}
\usepackage{setspace}
\usepackage{inconsolata}
\usepackage{libertine}
\usepackage{amsmath,amssymb,amsthm}
\usepackage{mathtools}
\usepackage{comment}
\usepackage[all,defaultlines=3]{nowidow}
\usepackage{microtype}
\usepackage{mathrsfs}
\usepackage{mathtools}
\usepackage{marvosym}
\usepackage{tikz}
\usepackage[export]{adjustbox}
\usepackage{todonotes}
\usepackage{forest}

\usetikzlibrary{calc}
\usetikzlibrary{shapes}
\definecolor{darkgreen}{rgb}{10,117,28}
\usepackage{diagbox}
\usepackage{stackengine}
\usepackage{mathrsfs}

\definecolor{blue}{rgb}{0.1,0.2,0.5}
\definecolor{brown}{rgb}{0.6,0.6,0.2}

\makeatletter
\def\ifenv#1{
   \def\@tempa{#1}%
   \ifx\@tempa\@currenvir
      \expandafter\@firstoftwo
    \else
      \expandafter\@secondoftwo
   \fi
}
\let\wfs@comment@comment\comment
\let\comment\@undefined
\newcommand{\untoto}{\let\toto\@undefined}
\usepackage{changes}
\let\wfs@changes@comment\comment
\let\comment\@undefined

\newcommand\comment{%
    \ifthenelse{\equal{\@currenvir}{comment}}
    {\wfs@comment@comment}
    {\wfs@changes@comment}%
}
\makeatother

\newtheorem{property}{Property}

\newcommand{\Cc}{\mathscr{C}}
\newcommand{\Ff}{\mathcal{F}}

\newcommand{\Qq}{\mathcal Q}

\renewcommand{\phi}{\varphi}

\newcommand{\Pp}{{\cal{P}}}
\newcommand{\N}{\mathbb{N}}

\newcommand{\bag}{\mathsf{bag}}
\newcommand{\sat}{\mathsf{sat}}
\newcommand{\cone}{\mathsf{cone}}
\newcommand{\cmp}{\mathsf{comp}}
\newcommand{\mrg}{\mathsf{mrg}}
\newcommand{\adh}{\mathsf{adh}}
\newcommand{\bgraph}{\mathsf{bgraph}}

\newcommand{\parent}{\mathsf{parent}}
\newcommand{\children}{\mathsf{children}}

\newcommand{\baggraph}{bag graph\xspace}
\newcommand{\baggraphs}{bag graphs\xspace}

\newcommand{\Oof}{\mathcal{O}}

\newcommand{\al}{\alpha}
\newcommand{\be}{\beta}
\newcommand{\DCD}{\textsc{Dominated Cluster Deletion}\xspace}
\newcommand{\ADCD}{\textsc{Annotated Dominated Cluster Deletion}\xspace}
\newcommand{\EDDC}{\textsc{Elimination Distance to Dominated Clusters}\xspace}
\newcommand{\AEDDC}{\textsc{Annotated Elimination Distance to Dominated Clusters}\xspace}
\newcommand{\APD}{\textsc{Annotated Partial Domination}\xspace}

\newcommand{\BW}{Black-White tree\xspace}
\renewcommand{\O}{\mathcal{O}}

\usepackage{accents}

\renewcommand{\succeq}{\succcurlyeq}

\usepackage{wasysym}
\usepackage{xspace}

\begin{document}

\maketitle

\begin{abstract}
  \noindent In the \textsc{Dominated Cluster Deletion} problem, we are given an undirected graph $G$ and integers $k$ and $d$ and the question is to decide whether there exists a set of at most $k$ vertices whose removal results in a graph in which each connected component has a dominating set of size at most $d$.
  In the \textsc{Elimination Distance to Dominated Clusters} problem, we are again given an undirected graph~$G$ and integers $k$ and $d$ and the question is to decide whether we can \emph{recursively} delete vertices up to depth $k$ such that each remaining connected component has a dominating set of size at most $d$.
  Bentert et al.~[Bentert et al., MFCS 2024] recently provided an almost complete classification of the parameterized complexity of \textsc{Dominated Cluster Deletion} with respect to the parameters $k$, $d$, $c$, and $\Delta$, where $c$ and $\Delta$ are the degeneracy, and the maximum degree of the input graph, respectively.
  In particular, they provided a non-uniform algorithm with running time $f(k,d)\cdot n^{\Oof(d)}$.
  They left as an open problem whether the problem is fixed-parameter tractable with respect to the parameter~$k + d + c$.
  We provide a uniform algorithm running in time $f(k,d)\cdot n^{\Oof(d)}$ for both \textsc{Dominated Cluster Deletion} and \textsc{Elimination Distance to Dominated Clusters}.
  We furthermore show that both problems are FPT when parameterized by~$k+d+\ell$, where $\ell$ is the semi-ladder index of the input graph, a parameter that is upper bounded and may be much smaller than the degeneracy $c$, positively answering the open question of Bentert~et~al.
  We further complete the picture by providing an almost full classification for the parameterized complexity and kernelization complexity of \textsc{Elimination Distance to Dominated Clusters}.
  The one difficult base case that remains open is whether \textsc{Treedepth} (the case $d=0$) is \textsf{NP-}hard on graphs of bounded maximum degree.
\end{abstract}

\section{Introduction}
Assume that $\mathcal{Q}$ is a graph problem that is hard to solve on general graphs but $\Cc$ is a class of graphs where~$\Qq$ can be solved efficiently.
In many cases, we can also solve $\Qq$ efficiently on instances that are \emph{close} to the instances of $\Cc$, say, on graphs that belong to $\Cc$ after the deletion of a small number of vertices.
Guo et al.~\cite{guo2004structural} formalized this concept under the name \emph{distance from triviality}.
For example, the size of a minimum vertex cover is the distance to the class of edgeless graphs and the size of a minimum feedback vertex set is the distance to the class of acyclic graphs. 
\textsc{Cluster Vertex Deletion} is the problem to determine for a given graph and number $k$ whether there exists a set of at most $k$ vertices whose deletion results in a \emph{cluster graph}, that is, a graph in which every connected component is a clique~\cite{bessy2023kernelization, tsur2021faster}. 
Many generalizations and special cases of \textsc{Cluster Vertex Deletion} have been studied.
For example, one may require that each connected component in the resulting graph is an $s$-club (a graph of diameter at most s) \cite{figiel20212, liu2012editing} or an $s$-plex (a graph in which each vertex has degree at least $n - s$) \cite{guo2010more,van2012approximation}.
Note that the special case where each clique in the solution graph has to be of
size one is again the \textsc{Vertex Cover} problem. 
Very recently, Bentert et al.~\cite{bentert2024breaking} studied the following \textsc{Dominated Cluster Deletion} problem: Given a graph~$G$ and integer parameters~$k$ and~$d$, does there exist a set $S$ of at most $k$ vertices such that each connected component of $G-S$ can be dominated by at most $d$ vertices? 
Bentert et al.\ provided an almost complete classification of the parameterized complexity and kernelization complexity of \textsc{Dominated Cluster Deletion} with respect to the parameters $k$, $d$, $c$, and $\Delta$, where $c$ and $\Delta$ are the degeneracy and the maximum degree of the input graph, respectively.
They proved the following results: \textsc{Dominated Cluster Deletion} is \textsf{para-NP}-hard for parameters $k+\Delta$ and $d+\Delta$, $W[2]$-hard for parameter $k+d$ and admits a non-uniform algorithm with running time $f(k,d)\cdot n^{\Oof(d)}$, and is non-uniformly fixed-parameter tractable with respect to parameter $k+d+\Delta$.
They left as an open problem whether the problem is fixed-parameter tractable with respect to the parameter $k + d + c$. 
They also showed that the problem does not admit a polynomial kernel even for $d=1$ 
with respect to the parameter $k + c$, or with respect to the parameter $k+d+\Delta$, unless $\textsf{NP}\subseteq \textsf{coNP/poly}$.

A related measure is the \emph{elimination distance} to a class $\Cc$, which measures the number of \emph{recursive} deletions of vertices needed for a graph $G$ to become a member of $\Cc$. 
Formally, a graph~$G$ has elimination distance $0$ to $\Cc$ if $G\in \Cc$, and otherwise elimination distance $k + 1$, if in every connected component of $G$, we can delete a vertex such that the resulting graph has elimination distance $k$ to~$\Cc$.
Note that elimination distance to the class containing only the empty graph corresponds to the well-known measure \emph{treedepth}.
Elimination distance was introduced by Bulian and Dawar~\cite{bulian2016graph} in their study of the parameterized complexity of the graph isomorphism problem. 
Just as distance to triviality, but with more general applicability, elimination distance allows to lift algorithmic results from a base class to more general graph classes. 
For example, Bulian and Dawar~\cite{bulian2016graph} provided an FPT algorithm for the graph isomorphism problem parameterized by the elimination distance to the class~$\Cc_d$ of graphs with maximum degree bounded by $d$, for any fixed integer $d$.
Hols et al.~\cite{hols2022elimination} proved the existence of polynomial kernels for the vertex cover problem parameterized by the size of a deletion set to graphs of bounded elimination distance to different classes of graphs. 
Agrawal et al.~\cite{agrawal2022fixed} proved fixed-parameter tractability of \textsc{Elimination Distance to $\Cc$} for any class $\Cc$ that can be defined by a finite set of forbidden induced subgraphs. 
Agrawal and Ramanujan studied the \textsc{Elimination Distance to Cluster Graphs}~\cite{agrawal2020parameterized}. 
Fomin, Golovach and Thilikos further generalized several of these prior results
by considering elimination distances to graph classes expressible by restricted first-order formulas~\cite{fomin2022parameterized}. 
These results are also implied by the recent algorithmic meta theorem for separator logic~\cite{PilipczukSSTV22,schirrmacher2023first}.
The related concept of recursive backdoors has recently been studied in
the context of efficient SAT solving~\cite{dreier2024sat,MahlmannSV21}. To the best of our knowledge, we are the first to study elimination distance to classes with small dominating sets.

\medskip
In this work, we answer the open questions of Bentert et al.~\cite{bentert2024breaking} for \textsc{Dominated Cluster Deletion} and furthermore provide an almost full classification of the parameterized complexity and kernelization complexity of \textsc{Elimination Distance to Dominated Clusters}.
More precisely, we prove that \textsc{Dominated Cluster Deletion} admits a uniform algorithm with running time $f(k,d)\cdot n^{\Oof(d)}$ as well as a uniform fixed-parameter algorithm with running time $f(k,d,\ell)\cdot n^{\Oof(1)}$, where $\ell$ is the semi-ladder index of the input graph. 
The semi-ladder index was introduced by Fabia\'nski et al.~\cite{FabianskiPST19arxiv} in the study of domination-type problems and is a parameter that is upper bounded and may be much smaller than the degeneracy~$c$ of a graph.
In fact, our result holds for all classes of graphs where the related \textsc{Partial Domination} problem is fixed-parameter tractable and satisfies an additional technical condition (the concept of \textsc{Partial Domination} and the details will be discussed in a moment).
In particular, our result answers the question of Bentert et al.\ positively.

Our main contributions are the uniform FPT algorithms for \DCD and \EDDC on semi-ladder-free graphs.
\begin{restatable}{theorem}{maintheoremDCD}\label{main-theorem-DCD}
  \DCD can be solved in time $f(k,d)\cdot n^{\Oof(d)}$ and in time \mbox{$f(k,d,\ell)\cdot n^{\Oof(1)}$} for a computable function $f$, where $\ell$ is the semi-ladder index of the input graph.
\end{restatable}
\begin{restatable}{theorem}{maintheoremEDDC}\label{main-theorem-EDDC}
  \EDDC can be solved in time $f(k,d)\cdot n^{\Oof(d)}$ and $f(k,d,\ell)\cdot n^{\Oof(1)}$ for a computable function $f$, where $\ell$ is the semi-ladder index of the input graph.
\end{restatable}

The hardness results completing the classification are simple observations. We prove the following.
\textsc{Elimination Distance to Dominated Clusters} is

\begin{enumerate}
    \item \textsf{para-NP}-hard for parameter $k+\Delta$; this is a simple consequence of the fact that the case $k=0$ corresponds to the \textsc{Dominating Set} problem, which is known to be \textsf{NP}-hard on graphs of maximum degree $3$~\cite{garey1979computers}, and
    \item \textsf{para-NP}-hard for parameter $d$; this is a simple consequence of the fact that the case $d=0$ corresponds to the \textsc{Treedepth} problem, which is known to be \textsf{NP}-hard~\cite{pothen1988complexity}, and
    \item $\textsf{W}[2]$-hard for parameter $k+d$; this is again a simple consequence of the fact that the case $k=0$ corresponds to \textsc{Dominating Set}, which is \textsf{W}[2]-hard with respect to parameter $d$~\cite{downey1995fixed}, and 
    \item uniformly fixed-parameter tractable for parameter $k+d+\ell$. This is our main contribution, we discuss the details below, and 
    \item uniformly fixed-parameter tractable with respect to $k$ when $d$ is considered to be a constant, that is, admits a uniform algorithm with running time $f(k,d)\cdot n^{\Oof(d)}$.
    \item The problem does not admit a polynomial kernel with respect to parameter $k$ even for fixed parameters $d=0$ and $c=2$ unless $\textsf{NP}\subseteq \textsf{coNP/poly}$; The parameter $d=0$ corresponds to \textsc{Treedepth}. We prove the simple observation that \textsc{Treedepth} is NP-complete on \mbox{$2$-de}\-generate graphs.
    It is then trivial using an AND-cross-composition to prove that \textsc{Treedepth} on \mbox{$2$-degenerate} graphs does not admit a polynomial kernel
    unless $\textsf{NP}\subseteq \textsf{coNP/poly}$.
\end{enumerate} 

The one case that we could not resolve is the parameter $d+\Delta$. 
The case $d=0$ corresponds to \textsc{Treedepth}, but we could not resolve whether the problem is \textsf{NP}-hard on graphs of bounded maximum degree.
We state the following two conjectures. 
\setcounter{theorem}{0}

\begin{conjecture}
    \textsc{Treedepth} is \textsf{NP}-hard on some class of graphs with bounded maximum degree.
\end{conjecture}

\begin{conjecture}
    \textsc{Treedepth} is \textsf{NP}-hard on planar graphs.
\end{conjecture}

The latter conjecture is related to the long-standing open question about the complexity of \textsc{Treewidth} on planar graphs. On the other hand, it is known since 1997 that \textsc{Treewidth} is \textsf{NP}-hard on graphs with maximum degree 9~\cite{bodlaender1997treewidth} and more recently on graphs with maximum degree 3~\cite{DBLP:conf/iwpec/BodlaenderBJKLM23}.

\subsection{Techniques}

Our main contribution is the uniform algorithm with respect to the parameter $k+d+\ell$ (the algorithm for $k+d$ is a simple modification of this algorithm).

We follow the approach of Bentert et al.~\cite{bentert2024breaking} to reduce to unbreakable graphs, that is, graphs that cannot be separated into multiple large connected components by small separators.
Let us give some formal definitions. 
A \emph{separation} in a graph $G$ is a pair of vertex subsets $A,B\subseteq V(G)$ such that $A\cup B=V(G)$ and there is no edge with one endpoint in $A\setminus B$ and the other in $B\setminus A$.
The order of the separation~$(A,B)$ is the size of its \emph{separator} $A\cap B$. 
For $q,k\in \N$, a set $X\subseteq V(G)$ is \emph{$(q,k)$-unbreakable} if for every separation~$(A,B)$ of~$G$ of order at most $k$, we have $|A\cap X|\leq q$ or $|B\cap X|\leq q.$
Intuitively speaking, $X$~is $(q,k)$-unbreakable if no separation of order $k$ can break $X$ in a balanced way: one of the sides must contain at most $q$ vertices of $X$.
For example, cliques and $k+1$-connected graphs are $(k,k)$-unbreakable, and square grids are $(\O(k^2),k)$-unbreakable.

Lokshtanov et al.~\cite{LokshtanovR0Z18} proved a powerful meta theorem for logically defined graph properties. 
Whenever a CMSO-property (monadic second-order logic with modulo counting predicates) can be solved efficiently on unbreakable graphs (for appropriately chosen parameters $k$ and $q$), then it can also be decided efficiently on general graphs.
There is a small caveat though. 
The meta theorem uses the non-constructive recursive understanding technique, and hence, one only obtains an efficient non-uniform algorithm on general graphs. 
Both \DCD and \EDDC are CMSO-properties, and hence fall into this framework. 
Bentert et al.\ rely on this framework and show how to solve \DCD on unbreakable graphs in time $f(k,d)\cdot n^{\Oof(d)}$. 
The key combinatorial observation is that in $(q,k)$-unbreakable graphs, the deletion of at most $k$ vertices leads to one giant component and a bounded number of remaining small components.
One can then guess a dominating set for the giant component in time $\Oof(n^d)$ and argue combinatorially to efficiently find the vertices to be deleted. 
Using the result of Lokshtanov et al.\ one obtains the non-uniform algorithm for general graphs with the desired running time. 

\begin{figure}
  \centering
  \includegraphics[width=.95\textwidth]{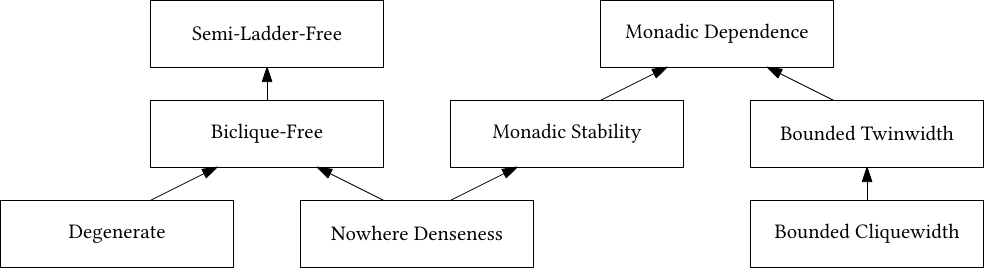}
  \caption{Relation between the discussed graph classes. Arrows indicate inclusion. \DCD and \EDDC are fixed-parameter tractable on monadically stable classes by a meta theorem for first-order logic. The same is true for bounded twinwidth classes when contraction sequences are given with the input. Our main result is the fixed-parameter tractability on semi-ladder-free graphs.}
  \label{fig:graph-classes}
\end{figure}

We remark that \DCD and \EDDC are in fact first-order definable problems, that is, for all parameters $k,d$, there exist first-order formulas of length $\O_{k,d}(1)$ defining the problems.
Hence, the problems are fixed-parameter tractable on all classes on which first-order model checking is fixed-parameter tractable, e.g.~on nowhere dense classes~\cite{grohe2017deciding} and even monadically stable graph classes~\cite{Dreier2024first,dreier2023first}, as well as on classes with bounded cliquewidth~\cite{courcelle2000linear} (and classes with bounded twinwidth~\cite{bonnet2021twin} when contraction sequences are given with the input).
Note that first-order logic cannot define connected components in general. However, the connected components arising in the recursive deletion of elements must have small diameter when they can be dominated by few vertices. Hence, connected components in positive instances of \textsc{Dominated Cluster Deletion} and \textsc{Elimination Distance to Dominated Clusters} become first-order definable.
Thus, on these classes one obtains algorithms with running time $f(k,d)\cdot n^{\Oof(1)}$ by the meta theorems for first-order logic.
It is conjectured that monadically dependent classes are the most general hereditary graph classes that admit efficient first-order model checking, we hence included them in the inclusion diagram, see~\cref{fig:graph-classes}.

This leaves two main questions. First, can we turn the non-uniform algorithm of Bentert et al.\ running in time $f(k,d)\cdot n^{\Oof(d)}$ into a uniform algorithm? 
Second, can we improve the running time on further restricted graph classes? 
We answer both questions positively. 

As mentioned, we follow the approach of Bentert et al.~\cite{bentert2024breaking} to reduce to unbreakable graphs. 
We first observe that on unbreakable graphs, the notion of elimination distance almost collapses to distance to triviality, enabling us to treat \DCD and \EDDC in a similar way. This idea has already been examined for hereditary classes in several papers see e.g.~\cite{AgrawalKLPRSZ22}. Note that being a dominated cluster is not a hereditary notion.
In order to avoid the non-uniform approach of Lokshtanov et al.~\cite{LokshtanovR0Z18}, we use a decomposition theorem due to Cygan et al.~\cite{CyganLPPS19}.
The theorem states that every graph admits a tree decomposition with small adhesion such that each bag is unbreakable in the subgraph induced by the bags below. 
We show that if an annotated version of the \textsc{Partial Dominating Set} problem can be solved in a slightly modified graph (that we call the bag graph), then we can implement an efficient dynamic programming algorithm along the tree decomposition to the original input graph for both \DCD and \EDDC.

The \textsc{(Annotated) Partial Dominating Set} problem is the following problem. 
We are given a graph with three sets $F$, $R$ and $B$ and parameters~$d$ and~$k$. 
We ask whether there exist $k$ vertices of~$G-F$ ($F$ stands for forbidden) that may be deleted such that the remaining graph has a Red-Blue dominating set of size at most $d$, i.e.~the vertices of $R$ can be dominated by $d$ vertices of $B$ ($R$~and~$B$ stand for Red and Blue).
Formally, the problem is defined as follows. 
Given a graph $G$ with (not necessarily disjoint) sets $F\subseteq V(G)$, $R\subseteq V(G)$, and $B\subseteq V(G)$ and parameters $d$ and~$k$,
does there exist 
a set $X\subseteq V(G)-F$ of size at most $k$ and
a set $D\subseteq B-X$ of size at most $d$
such that $G[R-X]$ is dominated by $D$?
The problem can obviously be solved in time $n^{\Oof(d)}$ (guess the dominating set and check if at most $k$ Red vertices remain undominated).
By our dynamic programming approach, we hence get a uniform algorithm running in time $f(k,d)\cdot n^{\Oof(d)}$ in general graphs.
We then turn our attention to restricted graph classes where \textsc{(Annotated) Partial Domination} can be solved more efficiently. 
Using an approach of Fabia\'nski et al.~\cite{FabianskiPST19arxiv}, we observe that \textsc{(Annotated) Partial Domination} can be solved efficiently on all \emph{semi-ladder-free} graph classes, which yields our second main result.

Semi-ladder-free are very general graph classes. For example, all degenerate classes are biclique-free (that is, exclude a fixed complete bipartite graph $K_{t,t}$ as a subgraph), which in turn are semi-ladder-free, hence, our result in particular answers the question of Bentert et al.\ positively.
On the other hand, these classes are incomparable with monadically stable classes and classes with bounded twinwidth on which we can efficiently solve the first-order model checking problem. 
For example, the class of all ladders has bounded twinwidth (in fact even bounded linear cliquewidth) but is not semi-ladder-free.
The class of all co-matchings is monadically stable (and has bounded linear cliquewidth as well) and is not semi-ladder-free.
On the other hand, the class of all $1$-subdivided cliques is semi-ladder-free (and even $2$-degenerate) but is neither monadically stable nor has bounded twinwidth. 

As mentioned, we require that \textsc{(Annotated) Partial Domination} needs to be solved on a slightly modified graph class (on the bag graphs), however, the property of being semi-ladder-free is preserved by this modification, so that ultimately we conclude our main results, \cref{main-theorem-DCD} and \cref{main-theorem-EDDC}.

\medskip\noindent\textbf{Organization.} The paper is organized as follows.
After recalling the necessary notation in \cref{sec:prelims}, 
we talk more precisely about partial domination and the graph parameters enabling FPT evaluation in~\cref{sec:partial-dom}.
We then present how to solve \DCD and \EDDC on unbreakable graph with bounded semi-ladder index in \cref{sec:unbreakable-bd-semi-ladder}.
After that, in \cref{sec:annotated}, we sketch the uniform algorithm on general semi-ladder-free graphs for \textsc{Dominated Cluster Deletion} and \textsc{Elimination Distance to Dominated Clusters} with running time $f(k,d)\cdot n^{\Oof(d)}$. The details are in the appendix.
The hardness results conclude the paper in \cref{sec:hard}.

\section{Preliminaries}\label{sec:prelims}

\textbf{Graphs.} 
We consider finite, undirected graphs
without loops. 
Our notation is standard, and we refer to Diestel's textbook for more background on graphs~\cite{diestel2024graph}.
We write $|G|$ for the size of the vertex set $V(G)$ and $\|G\|$ for $|V(G)|+|E(G)|$, where $E(G)$ is the edge set of a graph $G$.
A~connected component of $G$ is a maximal connected subgraph of $G$.
For a vertex subset $X\subseteq V(G)$, we write~$G[X]$ for the subgraph of $G$ induced by $X$.
We write $G-X$ for $G[V(G)\setminus X]$ and for singleton sets~$\{v\}$ we write $G-v$ instead of $G-\{v\}$. 
We denote the open neighborhood of a vertex~$v$ by~$N(v)$ and the closed neighborhood (including the vertex $v$) by $N[v]$.
For a set $Y\subseteq V(G)$, we write $N[Y]$ for $\bigcup_{v\in Y}N[y]$.

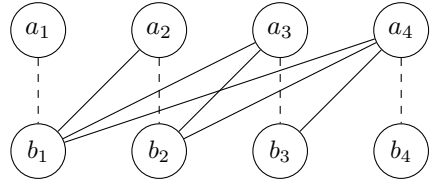
\begin{figure}[]
  \centering
  \begin{tikzpicture}[every node/.style={draw,circle},scale=.8]
    \node (a1) at (0,2) {$a_1$};
    \node (a2) at (2,2) {$a_2$};
    \node (a3) at (4,2) {$a_3$};
    \node (a4) at (6,2) {$a_4$};

    \node (b1) at (0,0) [scale=.95] {$b_1$};
    \node (b2) at (2,0) [scale=.95] {$b_2$};
    \node (b3) at (4,0) [scale=.95] {$b_3$};
    \node (b4) at (6,0) [scale=.95] {$b_4$};

    \path
    (a2) edge (b1)
    (a3) edge (b1)
    (a3) edge (b2)
    (a4) edge (b1)
    (a4) edge (b2)
    (a4) edge (b3);

    \path
    (a1) edge[dashed] (b1)
    (a2) edge[dashed] (b2)
    (a3) edge[dashed] (b3)
    (a4) edge[dashed] (b4);
  \end{tikzpicture}
  \caption{Semi-ladder of order 4. Solid lines represent edges while dashed edges represent non-edges.}
  \label{fig:semi-ladder}
\end{figure}

\medskip\noindent\textbf{Semi-ladders.}
Let $G$ be a graph. 
Two sequences, $a_1,\ldots,a_n\in V(G)$ and $b_1,\ldots,b_n\in V(G)$ of $2n$ distinct vertices, form a {\em{semi-ladder}} of order $n$ in $G$ if $\{a_i,b_j\}\in E(G)$ for all $i,j\in \{1,\ldots,n\}$ with $i>j$, and $\{a_i,b_i\}\notin E(G)$ for all $i\in \{1,\ldots,n\}$.
Note that we do not impose any condition for $i<j$.
The \emph{semi-ladder index} of a graph
is the maximum order of a semi-ladder that it contains.
A class of graphs has \emph{bounded semi-ladder index} or is called \emph{semi-ladder-free} if there is some $\ell$ such that the semi-ladder index of all of its members is bounded by $\ell$, see \cref{fig:semi-ladder}.

\medskip
\noindent \textbf{Tree decompositions.}
A \emph{tree decomposition} of a graph $G$ is a pair $\mathcal{T}=(T,\bag)$, where $T$ is a rooted tree and $\bag\colon V(T)\to 2^{V(G)}$ is a mapping that assigns to each node $x\in T$ its bag $\bag(x)\subseteq V(G)$ such that the following conditions are satisfied:
\begin{itemize}
  \item For every vertex $v\in V(G)$, there exists a node $x\in V(T)$ such that $v\in\bag(x)$.
  \item For every vertex $v\in V(G)$, the set of nodes $x\in V(T)$ satisfying $v\in\bag(x)$ induces a connected subtree of $T$.
  \item For every edge $\{u,v\}\in E(G)$, there exists a node $x\in V(T)$ such that $u,v\in\bag(x)$.
\end{itemize}

Let $\mathcal{T}=(T,\bag)$ be a tree decomposition of a graph $G$.
The \emph{adhesion} of a node $x\in V(T)$ is defined as $\adh(x)\coloneqq \bag(\parent(x))\cap \bag(x)$ and the \emph{margin} of a node $x\in V(T)$ is defined as $\mrg(x)\coloneqq \bag(x)\setminus \adh(x)$.
The \emph{cone} at a node $x\in V(T)$ is defined as $\cone(x)\coloneqq \bigcup_{y\succeq_T x} \bag(y)$ and the \emph{component} at a node $x\in V(T)$ is defined as $\cmp(x)\coloneqq \cone(x)\setminus \adh(x)= \bigcup_{y\succeq_T x} \mrg(y)$.
Here, $y\succeq_T x$ means that $y$ is a descendant of $x$ in $T$.

\medskip
\noindent\textbf{Unbreakable graphs.}
A \emph{separation} in a graph $G$ is a pair of vertex subsets $A,B\subseteq V(G)$ such that $A\cup B=V(G)$ and there is no edge with one endpoint in $A\setminus B$ and the other in $B\setminus A$.
The order of the separation~$(A,B)$ is the size of its \emph{separator} $A\cap B$.

For $q,k\in \N$, a vertex subset $X$ in a graph $G$ is \emph{$(q,k)$-unbreakable} if for every separation~$(A,B)$ of~$G$ of order at most $k$, we have $|A\cap X|\leq q$ or $|B\cap X|\leq q.$
Not every graph is $(q,k)$-unbreakable, but all graphs admit tree decompositions with small adhesion and unbreakable parts. 
For fixed~$q,k\in\N$, a tree decomposition $(T,\bag)$ of a graph $G$ is $(q,k)$-unbreakable if for every node $x\in V(T)$, the bag $\bag(x)$ is $(q,k)$-unbreakable in $G[\cone(x)]$.
By the result of Cygan et al.~\cite{CyganLPPS19}, such tree decompositions exist and can be computed efficiently.

\begin{theorem}[Theorem 10 of~\cite{CyganLPPS19}]
  \label{thm:CyganLPPS19}
  Let $G$ be a graph and $k$ an integer.
  There exists an integer $q\in 2^{\mathcal{O}(k^2)}$ and an algorithm that computes in time $q\cdot |G|^2\|G\|$ a tree decomposition $(T,\bag)$ of $G$ with at most $|G|$ nodes such that the following conditions hold:
  \begin{itemize}
    \item For every node $x\in V(T)$, the bag $\bag(x)$ is $(q,k)$-unbreakable in the subgraph~$G[\cone(x)]$.
    \item For every node $x\in V(T)$, the adhesion $\adh(x)$ is of size at most $q$.
  \end{itemize}
\end{theorem}

A tree decomposition $(T,\bag)$ is \emph{regular} if for every non-root node $x\in V(T)$,
\begin{itemize}
\item the margin $\mrg(x)$ is non-empty,
\item the graph $G[\cmp(x)]$ is connected, and
\item every vertex of $\adh(x)$ has a neighbor in $\cmp(x)$.
\end{itemize}

We may further assume that the tree decomposition we get from \cref{thm:CyganLPPS19} is regular, see the construction of Lemma 2.8 of~\cite{bojanczyk2016definability} and the discussion in~\cite{PilipczukSSTV22}.

\medskip
\noindent\textbf{Parameterized complexity.} 
A parameterized problem $L\subseteq \Sigma^*\times \N$ is \emph{fixed-parameter tractable (FPT)} if there exists an algorithm that decides for an instance $(x,k)$ whether $(x,k)\in L$ in time $f(k)\cdot |(x,k)|^{\Oof(1)}$ for some computable function $f$.

\section{Partial domination and FPT evaluation}\label{sec:partial-dom}

We show how to efficiently solve the \APD problem on semi-ladder-free graphs.
This problem falls into the framework of domination-type problems on semi-ladder-free graphs introduced by Fabia\'nski et al.~\cite{FabianskiPST19arxiv}.

We are going to use the following meta theorem of Fabia\'nski et al.~\cite{FabianskiPST19arxiv}. 
A \emph{domination-type problem} is a formula $\exists \bar x \forall \bar y \delta(\bar x,\bar y)$, where $\delta$ is a positive first-order formula using fixed distances as predicates (called a \emph{distance formula} in~\cite{FabianskiPST19arxiv}).
On every graph $G$, a formula $\delta(\bar x, \bar y)$ defines a bipartite graph $G_\delta$ on the vertex set $V(G)^{|\bar x|}\times V(G)^{|\bar y|}$, where two elements $\bar u$ and $\bar v$ are connected by an edge if the formula $\delta(\bar u,\bar v)$ is true in $G$.
The semi-ladder index of $\delta$ on $G$ is defined as the semi-ladder index of~$G_\delta$. 

\begin{theorem}
  [Theorem 19 of~\cite{FabianskiPST19arxiv}]
  \label{thm:partial-dom-algorithm}
  Let $\Cc$ be a class of graphs and let $\delta$ be a distance formula. Assume that $G_\delta$ has bounded semi-ladder index~$\ell$ for some fixed $\ell\in\mathbb{N}$ on all graphs $G\in \Cc$.
  Then there is an algorithm that solves the domination-type problem $\delta$ on graphs $G$ from $\Cc$ in time $f(\ell, |\delta|)\cdot \|G\|$.
\end{theorem}

\textsc{Annotated Partial Domination} can be written with the following first-order formula:
\begin{align*}
  \phi_{k,d}=\exists x_1\ldots \exists x_k & \exists y_1\ldots \exists y_d \forall z ~~
  \Big( \bigwedge_{1\leq i\leq k}\neg F(x_i)\Big) \wedge 
  \Big( \bigwedge_{1\leq i\leq d} B(y_i)\Big) \wedge \\
  &\Big( R(z) \rightarrow \Big(\bigvee_{1\leq i\leq k} \mathrm{dist}(z,x_i)=0\vee \bigvee_{1\leq i\leq d}\mathrm{dist}(z,y_i)\leq 1\Big) \Big)
\end{align*}

However, $\phi$ is not a distance formula as it uses negations. This can be circumvented by inverting the predicates, i.e.~$F' = V(G)-F$ and $R' = V(G)-R$. We then obtain the following formula
\begin{align*}
  \delta_{k,d}=\exists x_1\ldots \exists x_k & \exists y_1\ldots \exists y_d \forall z ~~
  \Big( \bigwedge_{1\leq i\leq k} F'(x_i)\Big) \wedge 
  \Big( \bigwedge_{1\leq i\leq d} B(y_i)\Big) \wedge \\
  &\Big( R'(z) \vee \bigvee_{1\leq i\leq k} \mathrm{dist}(z,x_i)=0\vee \bigvee_{1\leq i\leq d}\mathrm{dist}(z,y_i)\leq 1 \Big)
\end{align*}

We prove next that $\delta_{d,k}$ has bounded semi-ladder index on every graph $G$ with bounded semi-ladder index. 

\begin{lemma}
  Let $G$ be a graph with semi-ladder index $\ell$. Then $\delta_{d,k}$ has semi-ladder index at most $f(d,k,\ell)$ on $G$, for some computable function $f$. 
\end{lemma}
\begin{proof}
  Lemma~4 of~\cite{FabianskiPST19arxiv} states that if $\phi_1(\bar x;\bar y),\ldots,\phi_k(\bar x;\bar y)$ are formulas and $\psi(\bar x;\bar y)$ is a positive boolean combination   of $\phi_1,\ldots,\phi_k$,
  and $G$ is a graph such that $\phi_1(G),\ldots,\phi_k(G)$ have bounded semi-ladder index, then also~$\psi(G)$ has bounded semi-ladder index. 
  This is the case for the above formula $\delta_{k,d}$. 
  By Lemma~5 of~\cite{FabianskiPST19arxiv}, the formula $\bigvee_{1\leq i\leq d}\mathrm{dist}(z,y_i)\leq 1$ has bounded semi-ladder index on every graph with bounded semi-ladder index. All other formulas trivially have bounded semi-ladder index on all graphs. 
\end{proof}

\begin{corollary}\label{cor:partial-dom-fpt}
  Let $\Cc$ be a class of graphs with bounded semi-ladder index $\ell$. Then \textsc{Annotated Partial Domination} can be solved in time $f(d,k,\ell)\cdot \|G\|$ for all $G\in \Cc$ for a computable function $f$. 
\end{corollary}

While the semi-ladder index is currently one of the most general graph parameters enabling FPT evaluation of $\delta_{k,d}$, some other parameter could have this property.

\begin{property}\label{para-dom-set}
  A graph parameter $t(\cdot)$ has \cref{para-dom-set} if for every graph $G$ \textsc{Annotated Partial Domination} can be solved in time $f(d,k,t(G))\cdot |G|^{\Oof(1)}$ for a computable function $f$. 
\end{property}

As previously discussed, examples of parameters with \cref{para-dom-set} include treewidth, maximum degree, Hadwiger number (i.e.~largest clique minor), degeneracy, largest excluded bi-clique, as they all imply bounded semi-ladder index. 
Other parameters, which are however not as interesting, as the problem can be solved by efficient FO model checking, are cliquewidth and twinwidth (when contraction sequences are given with the input).

As mentioned, we want to solve \textsc{Annotated Partial Domination} on slightly modified graphs. See \cref{def:bgraph} for the precise definition of bag graphs. We therefore want a parameter that, if bounded on the graph, bounds a parameter with \cref{para-dom-set} on \baggraphs. 

\begin{property}\label{para-bound-bgraph}
  A graph parameter $t_2(\cdot)$ has \cref{para-bound-bgraph} if there is a parameter $t_1(\cdot)$ satisfying \cref{para-dom-set} such that for every graph $G$, and integers $q,k,d$, every $(q,k)$-unbreakable tree decomposition $(T,\bag)$, vertex $x\in T$, and set $S\subseteq\cone(x)$ we have that 
  $t_1(\bgraph_S(x)) \le f(q,k,d,t_2(G))$. 
\end{property}

In the rest of the paper, we precisely reference parameters having \cref{para-dom-set} or \cref{para-bound-bgraph}.
In particular, the semi-ladder index also satisfies \cref{para-bound-bgraph}, see~\cref{lem:para-bgraph}.

\section{Algorithm on unbreakable graphs based on \textsc{Annotated Partial Domination}}\label{sec:unbreakable-bd-semi-ladder}

As sketched above, we first solve \textsc{Dominated Cluster Deletion} and \textsc{Elimination Distance to Dominated Clusters} on unbreakable graphs. Our algorithm runs efficiently on all graph classes for which we can efficiently solve the \textsc{Annotated Partial Domination} problem, that is, for classes on which a parameter $t(\cdot)$ satisfying \cref{para-dom-set} is bounded. 

In the following, $q,k$, and $d$ will always be non-negative integers, and we will not always quantify them for readability.
We will also tacitly assume that our graphs have at least $2q+1$ vertices. 
Then, whenever we delete a set $S$ of at most $k$ vertices there will be a unique connected component~$C_0$ of $G-S$ with more than $q$ vertices that we call the \emph{large connected component}.
When a graph has less vertices, we can solve all our problems by brute-force.

\begin{lemma}\label{lem:remove-v-unbreakable}
  Let $G$ be a $(q,k)$-unbreakable graph, and let $v\in V(G)$. Then $G-v$ is $(q,k-1)$-unbreakable.
\end{lemma}
\begin{proof}
  Assume that $G-v$ that is not $(q,k-1)$-unbreakable.
  Then, there is a separation $(A,B)$ of $G-v$ such that $|A|\geq q$, $|B|\geq q$ and $|A\cap B|\leq k-1$.
  By adding the vertex $v$ to the separator, we obtain a separation $(A',B')$ of $G$ such that $|A'|>q$, $|B'|>q$ and $|A'\cap B'|\leq k$ contradicting that $G$ is $(q,k)$-unbreakable.
\end{proof}

\begin{lemma}\label{lem:neg-instance}
  Let $G$ be a $(q,k)$-unbreakable graph. If $G$ is a positive instance of \textsc{Dominated Cluster Deletion} with parameters $k$ and $d$, then $G$ has a dominating set of size at most $q+d$.
\end{lemma}

\begin{proof}
  Let $S'$ with $|S'|\leq k$ be a solution for \textsc{Dominated Cluster Deletion} with parameters~$d$ and~$k$. 
  Let $A=S'\cup C_0$, where $C_0$ is the large connected component of $G-S'$ and \linebreak$B=S'\cup \bigcup_{1\leq j\leq m}C_j$, where $C_1,\ldots, C_m$ are the other connected components of $G-S'$.
  Then, $(A,B)$ is a separation of order at most~$k$, hence $|B|\leq q$, as $G$ is $(q,k)$-unbreakable.
  As $G$ is a positive instance of \textsc{Dominated Cluster Deletion}, $C_0$ can be dominated by a set $D_0$ of size at most $d$. 
  Then $D_0\cup B$ is dominating set of $G$ of size at most $q+d$. 
\end{proof}
\subsection{\textsc{Dominated Cluster Deletion}}

We first deal with \textsc{Dominated Cluster Deletion}. 
Our approach is based on finding \emph{skeletons for dominated cluster deletion}, which are vertex sets that can be extended to the deletion set of a solution for \textsc{Dominated Cluster Deletion} and in $(q,k)$-unbreakable graphs capture the part of a solution that essentially separates the large connected component from the other connected components.

\begin{definition}
  \label{def:skeleton}
  Let $G$ be a $(q,k)$-unbreakable graph. 
  We call a set $S$ a \emph{skeleton for dominated cluster deletion} for parameters~$k$ and $d$ if there is a superset $S'\supseteq S$ of vertices satisfying:
  \begin{itemize}
    \item $|S'|\le k$, 
    \item every connected component of $G-S'$ has domination number at most $d$, and
    \item $S$ contains exactly the vertices of $S'$ that have at least one neighbor in $C'_0$ and at least one neighbor in \mbox{$G-N[C'_0]$},
    where $C'_0$ is the large connected component of $G-S'$.
  \end{itemize}
\end{definition}

\begin{figure}
	\centering
	\begin{subfigure}{0.45\textwidth}
	  \includegraphics[page=1, width=\textwidth]{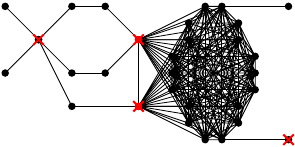}
	\end{subfigure}
	\hfill
	\begin{subfigure}{0.45\textwidth}
	  \includegraphics[page=2, width=\textwidth]{figures/example2-skeleton.pdf}
	\end{subfigure}
	\hfill
	\begin{subfigure}{0.45\textwidth}
	  \includegraphics[page=4, width=\textwidth]{figures/example2-skeleton.pdf}
	\end{subfigure}
	\hfill
	\begin{subfigure}{0.45\textwidth}
	  \includegraphics[page=5, width=\textwidth]{figures/example2-skeleton.pdf}
	\end{subfigure}
	\hfill
	\caption{Different skeletons for the dominated cluster deletion problem for parameter $k=4$ and $d=1$ on a fixed (14,4)-unbreakable graph. The skeleton $S$ is in red, the witness $S'$ also contains the crossed-out vertices.}
	\label{fig:example-skeleton}
\end{figure}

Note that $S$ may be the empty set in the third item above if $G-S'$ has only one connected component.
\cref{fig:example-skeleton} shows examples of skeletons for dominated cluster deletion.

The next lemma is the key ingredient to compute the set of skeletons (and to show that there are only few possible skeletons). It shows that there is a small set of vertices that hits every skeleton.

\begin{lemma}
  \label{lem:skeleton-dom-set}
  Let $q,k,d$ be integers, $G$ a $(q,k)$-unbreakable graph, and assume that $S$ is a non-empty skeleton for dominated cluster deletion for parameters $k$ and $d$.
  Given a dominating set $X$ of size at most $q+d$, there is a vertex $v\in S$ that is
  \begin{enumerate}
    \item in the dominating set $X$, or
    \item in the neighborhood $N(x)$ of a vertex $x\in X$, where $\deg(x)\le q$, or
    \item in the neighborhood $N(y)$ of a vertex $y\in N(x)$, where $\deg(x)\leq q$, $\deg(y)\leq q$, and $x\in X$.
  \end{enumerate}
\end{lemma}

\begin{proof}
  Assume that Case 1 is not satisfied so that no vertices from the skeleton are in the dominating set $X$.
  Denote by $C'_0$ the large connected component of $G-S'$ for a set $S'\supseteq S$ witnessing that $S$ is a skeleton, hence, $S$ contains exactly the vertices of $S'$ that have a neighbor in~$C'_0$ and a second neighbor in \mbox{$G-N[C'_0]$}.
  Let $C_0$ be the large connected component of $G-S$. Note that it contains~$C'_0$.

  Let $A=S\cup C_0$ and $B=S\cup \bigcup_{1\leq j\leq m}C_j$, where $C_1,\ldots, C_m$ are the other connected components of~$G-S$. Then $(A,B)$ is a separation of order at most $|S|\le k$, hence $|B|\leq q$, as~$G$ is $(q,k)$-unbreakable.
  Pick an arbitrary element $v\in S$. By definition of $S$, $v$ has a neighbor in $C_0$ and a neighbor $w$ in a second connected component, say $C_1$ of $G-S$. Hence, $w$ must be a vertex of degree at most $q$ as its neighborhood lies in $G-C_0\subseteq B$.
  If $w$ is in $X$, then this vertex $v$ satisfies Case~2. Otherwise, $w$ must have a neighbor $x$ in $X$ (since $X$ is a dominating set). Since we are not in Case~1, this $x$ is not in $S$, and is therefore in $C_1$. Hence, similarly to $w$, we have that $x$ has degree at most~$q$, and we are in Case 3.
\end{proof}

\cref{lem:skeleton-dom-set} gives rise to a bounded search tree branching algorithm that branches over the possible choices for a vertex from $S$ given by the conditions of the lemma.

\begin{definition}
  The {\em skeleton-algorithm} with parameter $q,k,d$ over a graph $G$ returns a family of sets of vertices, each set of size at most $k$.
  First, it computes a dominating set $X$ of size at most $q+d$ (returning the empty family if there is no such set).
  Second, it computes the set $X\cup Y\cup Z$,
  where $Y$ is the set of all vertices in the neighborhood $N(x)$ of a vertex $x\in X$ with $\deg(x)\leq q$, 
  and the set $Z$ is the set of all vertices in the neighborhood $N(y)$ of a vertex $y\in Y$  where $\deg(y)\leq q$.
  Third, call the algorithm with parameters $q,k-1,d$ on $G-v$ for every $v$ in $X\cup Y\cup Z$ (and get a family $\Ff_v$).
  Finally, for every $v$ in $X\cup Y\cup Z$ add $v$ to every set in $\Ff_v$, and return the empty set and the union of all the families $\Ff_v$. 
\end{definition}

\begin{algorithm}[h]
	\caption{Skeleton-algorithm}
	\begin{algorithmic}[1]
	  \Procedure{Skeleton}{$G,q,k,d$}
		\State $\Ff\coloneqq \emptyset$
		\If{$k=0$}
		  \State \Return $\Ff$
		\EndIf
		\State $X\coloneqq$ \Call{DominatingSet}{$G,q+d$} 
		\State $Y\coloneqq \{N(x)\colon x\in X\wedge\deg(x)\leq q\}$
		\State $Z\coloneqq \{N(y)\colon y\in Y\wedge\deg(y)\leq q\}$
		\ForAll{$v\in X\cup Y\cup Z$}
    \State $\Ff\coloneqq \Ff\cup \{\emptyset\} \cup \big(\{v\}\times$ \Call{Skeleton}{$G-v,q,k-1,d$}$\big)$
		\EndFor
    \State \Return $\Ff$
	  \EndProcedure
	\end{algorithmic}
  \label{alg:skeleton}
\end{algorithm}

Note that in the above definition, if a family $\Ff_v$ is empty, adding $v$ to every set still results in an empty family (e.g.~in the case where $G-v$ does not have a $q+d$ dominating set). Do not be confused with $\Ff_v$ having the empty set as one of the family members (e.g.~for branches of the algorithm that stops before reaching $k=0$).

We now show that the skeleton algorithm 1) runs efficiently and outputs a small family of sets which is proved in~\cref{lem:skeleton-quick}, and 2) if the given graph was a positive instance to \DCD, then the skeleton algorithm outputs all possible skeletons; see~\cref{lem:skeleton-correct}.
\begin{lemma}\label{lem:skeleton-quick}
  Let $t(\cdot)$ be a graph parameter satisfying \cref{para-dom-set}, then the skeleton-algorithm runs in time $f(q,k,d,t(G))\cdot |G|^{\Oof(1)}$, and outputs a family of at most $g(q,k,d)$ sets where $f$ and $g$ are computable functions.
\end{lemma}
\begin{proof}
  Computing $X,Y$ and $Z$ takes time $f(q,k,d,t(G))\cdot |G|^{\Oof(1)}$ thanks to the fact that $t(\cdot)$ satisfies~\cref{para-dom-set}.
  We then branch over $(q+d)q^2$ many choices for $v$, and the recursion has depth at most $k$.
\end{proof}

\begin{lemma}\label{lem:skeleton-correct}
  If $G$ is a $(q,k)$-unbreakable graph that is a positive instance to \DCD with parameters $(k,d)$, then the skeleton-algorithm outputs a family that contains every skeleton.
\end{lemma}
\begin{proof}
  By induction on $k$. If $k=0$, we simply output the empty set.
  Otherwise, by \cref{lem:neg-instance} there is such a set $X$, and by~\cref{lem:skeleton-dom-set} at least one element $v$ of $X$ is in $S$. For the correct choices of $v$, $G-v$ is a positive instance for \DCD with parameter $k-1,d$, and by \cref{lem:remove-v-unbreakable}, $G-v$ is $(q,k-1)$-unbreakable, so the recursive call to \textsc{Skeleton} produces the expected sets by induction. 
\end{proof}

It remains to verify that $S$ can indeed be enlarged to a set $S'$ of size at most $k$ with the properties.

\begin{lemma}\label{lem:skeleton-check}
  Given a $(q,k)$-unbreakable graph $G$ and a set $S$ of at most $k$ vertices, we can test in time $f(q,k,d,t(G))\cdot |G|^{\Oof(1)}$, whether $S$ is a skeleton for \DCD with parameters $k$ and $d$ assuming $t(\cdot)$ satisfies \cref{para-dom-set}.
\end{lemma}
\begin{proof}
  We need to check the existence of a set $S'\supseteq S$ with $|S'|\le k$ such that:
  \begin{itemize}
    \item every connected component of $G-S'$ has domination number at most $d$, and
    \item $S$ contains exactly the vertices of $S'$ that have a neighbor in $C'_0$ and a second neighbor in \mbox{$G-N[C'_0]$} where $C'_0$ is the large connected component of $G-S'$.
  \end{itemize}

  Note that by these conditions the large connected component $C_0$ of $G-S$ is separated from the small connected components by $S$. Let $s=|S|$, what remains to be done is hence the following. 

  \begin{enumerate}
    \item Find an optimal solution for $G-S-C_0$.
    This can be done in time $f(k,d)$, because this subgraph has size at most $q$ and yields a value $k'$ of vertices that need to be deleted.
    If $k'>k-s$, we conclude that this $S$ is not a skeleton because there is no possible $S'$.
    \item Test whether an additional set $W$ of size $k''=k-s-k'$ can be deleted from $C_0$ such that $G[C_0-W]$ can be dominated by $d$ vertices. Note that by assumption on $S$ and $S'$ the deletion of~$W$ will not cause $C_0$ to break into multiple further connected components.
    Hence, we simply check whether $C_0$ is a positive instance of \textsc{Annotated Partial Domination} with parameter~$k''$, where $R=B=C_0$, and $F=\emptyset$. If $S$ is a skeleton, such a set exists (as $S'\cap C_0$ is a candidate) and hence, if $G$ is a positive instance of \textsc{Dominated Cluster Deletion} a positive solution will be found.
    Since $t(\cdot)$ is a graph parameter satisfying \cref{para-dom-set}, we can test the existence of such~$W$ in time $f(k,d,t(G))\cdot|G|^{\Oof(1)}$. If no solution is found, then this $S$ is not a skeleton.\qedhere
  \end{enumerate}
\end{proof}

The combination of~\cref{lem:skeleton-dom-set,lem:skeleton-quick,lem:skeleton-correct,lem:skeleton-check} yields the following theorem.

\begin{theorem}\label{thm:algo-simple-unbreak}
  \textsc{Dominated Cluster Deletion} on $(q,k)$-unbreakable graphs can be solved in time $f(q,k,d,t(G))\cdot |G|^{\Oof(1)}$, where $t(\cdot)$ is a parameter satisfying \cref{para-dom-set}.
\end{theorem}

\subsection{\textsc{Elimination Distance to Dominated Clusters}}

We now show how to adapt the algorithm for \textsc{Elimination Distance to Dominated Clusters}. The approach is very similar. Of course, the notion of skeletons has to be adapted, however, they behave very similarly on unbreakable graphs. 

First, let us properly define \EDDC.
\begin{definition}
  A graph $G$ has {\em elimination distance} $k$ to a class $\Cc$ if:
  \begin{itemize}
    \item $G$ is not connected and every connected component of $G$ has elimination distance $k$ to $\Cc$, or
    \item $G$ is connected and there is a vertex $v$ such that $G-v$ has elimination distance $k-1$ to $\Cc$, or
    \item $G\in \Cc$ and $k=0$.
  \end{itemize}
\end{definition}

In our case we measure, for every $d$, the elimination distance to the class $\Cc_d$ of $d$-dominated clusters, i.e.~graphs where every connected component has a $d$-dominating set. We interchangeably write ``elimination distance $k$ to $d$-dominated cluster'', and ``a positive instance for $(k,d)$-\EDDC''.
For algorithmic purposes, the above definition is not the most useful. We will instead look at the set of vertices that needs to be deleted to get $d$-dominated clusters. While this set is of unbounded size, it has a nice tree structure that is much more useful for our algorithms. This characterization is pretty standard, see the definition of {\em elimination order}~\cite[Proposition 4.3]{bulian2016graph}

For a rooted tree $T$, we write $\preccurlyeq_T$ for the ancestor relation, that is, $x \preccurlyeq_T y$ if $x$ lies on the unique path from $y$ to the root of $T$.
Note that we treat every node as an ancestor of itself.
The least common ancestor of two nodes $y,z\in V(T)$ is the $\preccurlyeq_T$-maximal element $x$ with $x\preccurlyeq_T y$ and~$x \preccurlyeq_T z$.

\begin{definition}\label{def:tree-structure}
  Let $G$ be a graph and $S\subseteq V(G)$. 
  We say that $S$ is \emph{tree-structured} if there is a rooted tree $T$ and a bijection $\lambda:S\rightarrow V(T)$ such that every path in~$G$ between two vertices $u,v$ of~$S$ must contain a common ancestor in $T$ of $\lambda(u)$ and~$\lambda(v)$.
  The tree $T$ is called an \emph{elimination tree for~$S$}. 
  The \emph{elimination depth} of a tree-structured set with respect to the tree $T$ is the depth of $T$.\linebreak
  The elimination depth of $S$ is the minimum elimination depth of $S$ over all elimination trees for $S$. 
\end{definition}

\begin{remark}
  A graph $G$ is a positive instance for $(k,d)$-\EDDC if and only if there is a tree-structured set $S$ with elimination depth $k$ such that every connected component of $G-S$ has a $d$-dominating set. 
\end{remark}

\begin{lemma}\label{lem:components-tree-structured}
  Let $G$ be a $(q,k)$-unbreakable graph with at least $3q(k+q)$ vertices. Let $S$ be a tree-structured set with an elimination tree $T$ of depth at most $k$. Then 
  \begin{enumerate}
  \item the largest connected component $C_0$ of $G-S$ is uniquely determined, and 
  \item for every component $C$ of $G-S$ there is a subset $S(C)\subseteq S$ of size at most $k$ that separates $C$ from the rest, that is, $G-S(C)$ has $C$ as one of its components. 
  \end{enumerate}
\end{lemma}
\begin{proof}
  Let $C$ be a connected component of $G-S$.
  As $S$ is tree-structured by $T$, $C$ is not adjacent to two $\preccurlyeq_T$-incomparable elements, that is, the set $S(C)$ of neighbors of $C$ in $S$ consists of $\preccurlyeq_T$-comparable elements.
  $C$ is a connected component both of $G-S$ and of $G-S(C)$, as all neighbors of $C$ in $S$ are collected in $S(C)$.
  
  Assume now that there are two connected components $C_0,C_1$ of $G-S$ of size greater than~$q$. However, $C_0$ and $C_1$ are separated by $S(C_0)$ of size at most $k$, which contradicts the $(q,k)$-unbreakability of $G$. So there can only be one connected component of size greater than $q$. 
  
  We now show that there must be at least one such component.
  First note that there cannot be more than $3q$ connected component in $G-S$. Otherwise, there would be a balance separator of $T$ of size $1$, by deleting this vertex and its ancestors, we delete at most $k$ vertices and obtain a separation with both sides containing at least $q$ components of $G-S$, contradicting the unbreakability of~$G$. Last, as there are only $3q$ components $C$ and $|S(C)|\le k$ we have that $|S|\le 3qk$ hence $|G-S|\ge 3q^2$ so on of the component must contain at least $q$ vertices.
\end{proof}

\begin{lemma}[Analog of \cref{lem:neg-instance}]\label{lem:neg-instance-eddc}
  Let $G$ be a $(q,k)$-unbreakable graph. If $G$ is a positive instance of \EDDC with parameters $k$ and $d$, then $G$ has a dominating set of size at most $q+d$.
\end{lemma}
The proof can be taken verbatim from \cref{lem:neg-instance} with $S'$ replaced by $S'(C_0)$.

\begin{definition}
  \label{def:skeleton2}
  Let $G$ be a $(q,k)$-unbreakable graph. We call a set $S$ a \emph{skeleton for elimination distance to dominated clusters} for parameters~$k$ and $d$ if there is a tree-structured superset $S'\supseteq S$ of elimination depth at most $k$ satisfying:
  \begin{itemize}
    \item every connected component of $G-S'$ has domination number at most $d$, and
    \item $S$ contains exactly the vertices of $S'$ that have a neighbor in $C'_0$ and a second neighbor in $G-N[C'_0]$ where $C'_0$ is the large connected component of $G-S'$ (the next lemma shows that this connected component is uniquely determined even though $S'$ can be larger than $k$).
  \end{itemize}
\end{definition}

The following observation follows immediately from \cref{lem:components-tree-structured}. 
\begin{observation}
  Every skeleton has size at most $k$. 
\end{observation}

\begin{figure}
  \centering
  \includegraphics[page=1, width=.7\textwidth]{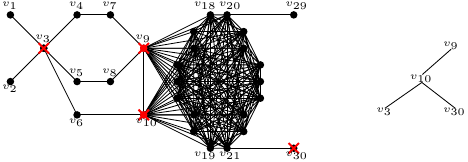}
  \caption{Example of a skeleton for elimination distance to dominated clusters for parameter $k=3$ and $d=1$ on a (12,3)-unbreakable graph (left). The skeleton $S$ is in red, the witness $S'$ also contains the crossed-out vertices. A tree structure for $S'$ of elimination depth $3$ is on the right.}
  \label{fig:example-skeleton-eddc}
\end{figure}

Now the analog of \cref{lem:skeleton-dom-set} also holds for skeletons for elimination distance to dominated clusters. The proof can be taken verbatim with $S'$ replaced by $S'(C_0)$.

\begin{lemma}
  \label{lem:skeleton-eddc}
  Let $q,k,d$ be integers, $G$ be a $(q,k)$-unbreakable graph and assume that $S$ is a non-empty skeleton for \EDDC with parameters $k$ and $d$.
  Given a domination set $X$ of size $q+d$, there is a vertex $v\in S$ that is
  \begin{enumerate}
    \item in the dominating set $X$, or
    \item in the neighborhood $N(x)$ of a vertex $x\in X$ where $\deg(x)\leq q$, or
    \item in the neighborhood $N(y)$ of a vertex $y\in N(x)$ with $\deg(x)\leq q$, $\deg(y)\leq q$, and $x\in X$.
  \end{enumerate}
\end{lemma}

Hence, the algorithm to guess a skeleton $S$ can be carried out in exactly the same way as for \textsc{Dominated Cluster Deletion}. 
We then guess a tree structure on $S$ (which takes time at most~$\Oof(k!)$). Note that these vertices being adjacent to the large connected component, the tree structure is a linear order.
In the final step, when we aim to extend a fixed skeleton $S$ to a tree-structured solution $S'$, the only difference is that we have to solve the base problem \textsc{Elimination Distance to Dominated Clusters} for the small part $G-S-C'_0$, which again can be simply solved by brute-force.
Note that because $G-S-C'_0$ is small, we can also find an elimination tree~$T$ for $S'$ and embed $S$ according to the originally guessed order.
We conclude the main theorem of this section. The full algorithm for the annotated version is presented in the appendix.

\begin{theorem}\label{thm:algo-simple-unbreak-eddc}
  \textsc{Elimination Distance to Dominated Clusters} on $(q,k)$-unbreakable graphs can be solved in time $f(q,k,d,t(G))\cdot |G|^{\Oof(1)}$, where $t(\cdot)$ is a parameter satisfying \cref{para-dom-set}.
\end{theorem}

\subsection{Skeletons for general graphs}
In this section, we show that by parameterizing only by $q,k$, and $d$, we can compute the skeleton of unbreakable graphs in time $f(q,k,d)\cdot |G|^{\O(d)}$. This construction follows the work of Bentert~et~al.~\cite{bentert2024breaking}. Having access to this algorithm will then be useful for our uniform version of their result.

\begin{lemma}\label{lem:poly-comput-skeleton}
  There is an algorithm that, given a $(q,k)$-unbreakable graph $G$, computes all skeletons~$S$ for dominated cluster deletion and for elimination distance to dominated clusters for parameters $k$ and~$d$ in time $f(q,k,d)\cdot n^{\O(d)}$.
\end{lemma}
This lemma resembles and can be proved analogously to \cite[Lemma 7]{bentert2024breaking}.
\begin{proof}
  Assume the existence of a solution $S'$ and let $S$ be its skeleton.
  We guess a set $D$ of at most $d$ vertices in time $O(n^d)$. (This will play the role of the dominating set for the big component). We then define $S_1 = G-N[D]$ and $S_2=\{v\in N(u)~|~u\in S_1 \wedge \deg(u)\le q\}$. If either $|S_1|>q$ or $|S_2|>q+kq$, we conclude that this $D$ is not a good candidate for the dominating set of the large connected component and terminate this branch. Otherwise, we now claim that $S\subseteq S_2$.

  To this end, let $C'_0$ be the large connected component of $G-S'$ and assume that the algorithm guessed~$D$ as a dominating set for $C'_0$.
  Observe that $C'_0\subseteq N[D] \subseteq N[C'_0]$. 
  A vertex of $G-N[C'_0]$ has no neighbor in $C'_0$, and therefore only has degree at most $q$.
  Finally, remember that any vertex~$v$ of~$S$ must have a neighbor $u$ that is not in $N[C'_0]$ and hence not in $N[D]$. So $u$ is in $S_1$ and $u$ has degree at most $q$. So $v$ belongs to $S_2$.

  Regarding the cardinality constraints. Note that at most $q$ vertices are in $G-C'_0$ and hence at most $q$ are in $G-N[D] = S_1$. Also, the only vertices of $S_2$ that are in $C_0$ must be neighbors of a vertex of $S'$ of degree at most $q$, hence $|S_2\cap C_0|\le kq$, and since $|G-C_0|\le q$, we get $|S_2|\le q+kq$.
  
  We conclude that $S\subseteq S_2$. So we can guess $S$ in time $f(q,k,d)\cdot n^{\O(d)}$.
\end{proof}

Note that \cref{lem:poly-comput-skeleton} applies to skeletons for both \DCD and \EDDC. Furthermore, the dominating set $D$ for the large connected component is computed (by brute-force).

\section{Annotated versions for \DCD}\label{sec:annotated}

\begin{definition}[\ADCD]
	Given a graph $G$, integers $k$, $d$, and vertex sets $F,R,B$, the $(k,d)$-\ADCD problem is to find at most $k$ vertices $S\subseteq G-F$ such that every connected component $C$ of $G-S$ has a Red-Blue dominating set of size at most $d$ (i.e.~$d$ Blue vertices dominating all Red vertices).
\end{definition}

The next proposition states that this problem can be solved on instances we call \baggraphs.

\begin{definition}\label{def:baggraphs}
	A graph $G$ is called a {\em $(q,k,d)$-\baggraph}, if the vertices of $G$ can be partitioned into two vertex sets $I,E$ called respectively {\em interior vertices} and {\em exterior vertices}, such that:
	\begin{itemize}
		\item every connected component of $G[E]$ has at most $2d+1$ vertices,
		\item every connected component of $G[E]$ has at most $q$ neighbors in $I$, and
		\item for every separation $L,R$ of $G$ of order $k$ such that $L\cap R \subseteq I$, we have that $|L\cap I|\le q$ or $|R\cap I|\le q$.
	\end{itemize}
\end{definition}

\begin{restatable}{proposition}{DcdBag}\label{prop:dcd-bag}
	There is an algorithm such that, given a $(q,k,d)$-\baggraph $G=(I,E)$, and vertex sets~$F,R,B$ of $G$ satisfying $E\subseteq F$ and $|E\cap B|\le qd$, the algorithm decides whether $G$ is a positive instance to $(k,d)$-\ADCD, in time $f(q,k,d,t(G))\cdot |G|^{O(1)}$, where~$t(\cdot)$ is a parameter satisfying~\cref{para-dom-set}.
\end{restatable}

We prove \cref{prop:dcd-bag} in \cref{sec:proof-dcd-bags}.
Using this, we then prove the following statement via dynamic programming on an unbreakable tree decomposition in~\cref{sec:dcd-tree-decomposition}.

\begin{theorem}\label{thm:CDCD}
	There is an algorithm solving \ADCD in time~$f(q,k,d,t(G))\cdot |G|^{O(1)}$, where $t(\cdot)$ is a parameter satisfying~\cref{para-bound-bgraph}.
\end{theorem}

Remember that \cref{para-bound-bgraph} implies that some parameter with \cref{para-dom-set} is bounded on bag graphs built in our algorithm, so we can use \cref{prop:dcd-bag} on these \baggraphs.
Note that \cref{thm:CDCD} implies \cref{main-theorem-DCD}, by setting $F=\emptyset$ and $R=B=V(G)$.
The annotated version for \EDDC is introduced and solved in~\cref{sec:aeddc}.

\section{Hardness Results}\label{sec:hard}

In this section, we give the details of the hardness results mentioned in the introduction.

\medskip\noindent{\bf Para-NP-hardness for parameter $k+\Delta$ and for parameter $d$}

A problem is \textsf{para-NP}-hard with respect to a parameter if the restriction to instances whose parameter value is at most a certain constant, is itself an \textsf{NP}-hard problem.

\begin{theorem}
    \textsc{Elimination Distance to Dominated Clusters} is \textsf{para-NP}-hard for parameter~$k+\Delta$. 
\end{theorem}
\begin{proof}
    Setting $k=0$ and $\Delta=3$ corresponds to \textsc{Dominating Set} on degree $3$ graphs. This problem is known to be \textsf{NP}-hard~\cite{garey1979computers}, which implies the statement. 
\end{proof}

\begin{theorem}
    \textsc{Elimination Distance to Dominated Clusters} is \textsf{para-NP}-hard for parameter $d$. 
\end{theorem}
\begin{proof}
    Setting $d=0$ corresponds to \textsc{Treedepth}. This problem is known to be \textsf{NP}-hard~\cite{pothen1988complexity}, which implies the statement.
\end{proof}

\medskip\noindent{\bf W[2]-hardness for parameter $k+d$}

A problem is \textsf{W}[2]-hard if a \textsf{W}[2]-hard problem reduces to it via an FPT-reduction.

\begin{theorem}
    \textsc{Elimination Distance to Dominated Clusters} is \textsf{W}[2]-hard for parameter $k+d$. 
\end{theorem}
\begin{proof}
    The \textsf{W}[2]-hard \textsc{Dominating Set}~\cite{downey1995fixed} reduces to \textsc{Elimination Distance to Dominated Clusters} by simply setting $k=0$. 
    This is an FPT-reduction.
\end{proof}

\medskip\noindent{\bf No polynomial kernels for parameter $k+c+d$}

A kernelization algorithm (or kernel) for a parameterized problem is a polynomial-time algorithm that maps each instance $(x,k)$ of a parameterized problem $L$ to an instance $(x',k')$ of $L$ such that 
\begin{itemize}
    \item $(x,k)\in L\Leftrightarrow (x',k')\in L$, and
    \item $|x'|+k'\leq f(k)$ for a computable function $f$. 
\end{itemize}

A kernel is polynomial if $f$ is. 
We use the AND-cross-composition technique~\cite{drucker2015new} to show that \textsc{Elimination Distance to Dominated Clusters} does not admit a polynomial kernel unless $\textsf{NP}\subseteq \textsf{coNP/poly}$. 
If an \textsf{NP}-hard language $L$ AND-cross-composes into a parameterized language~$Q$, then $Q$ does not admit a polynomial kernel, unless $\textsf{NP}\subseteq \textsf{coNP/poly}$.
Since the treedepth of a disjoint union of a family of graphs is equal to the maximum over the treedepth of these graphs, the disjoint union yields an AND-cross-composition from the unparameterized version of \textsc{Treedepth} into the parameterized one.
As computing the treedepth of a graph is \textsf{NP}-hard~\cite{pothen1988complexity}, it follows that \textsc{Treedepth} not admit a polynomial kernel, unless $\textsf{NP}\subseteq \textsf{coNP/poly}$.
It remains to prove that \textsc{Treedepth} is also \textsf{NP}-complete on $2$-degenerate graphs.
Recall that $c$ denotes the degeneracy of a graph. A graph $G$ is $c$-degenerate if every subgraph $H$ has a vertex of degree at most $c$ (in $H$). 

\begin{theorem}
    \textsc{Treedepth} is \textsf{NP}-hard on $2$-degenerate graphs.
\end{theorem}
\begin{proof}
    We reduce \textsc{Treedepth} to \textsc{Treedepth} on $2$-degenerate graphs. Given a graph $G$ and parameter $k$, we compute $H$ by replacing each edge of $G$ by two disjoint paths of length $2$ with the same endpoints. We set $k'=k+1$ and prove that $G$ has treedepth at most $k$ if and only if $H$ has treedepth at most $k'$.
    Observe that $H$ is $2$-degenerate. 

    Assume that $G$ has treedepth at most $k$ with an elimination tree $T$. 
    We take the same elimination tree for $H$ and in the very last step eliminate the isolated subdivision vertices. 

    Vice versa assume that $H$ has treedepth at most $k+1$ with an elimination tree of depth at most~$k+1$. 
    Observe that every subdivision vertex must be comparable in the tree-order with both of its endpoints, as each edge must be controlled by the ancestor-descendant relation in the elimination tree. 
    Now, if a sudivision vertex is eliminated before one of its endpoints, we can simply exchange the subdivision vertex with this endpoint and thereby obtain an elimination tree that is not deeper than the original one. 
    In this way, we obtain an elimination tree where the subdivision vertices are on the lowest level. Then the tree with the subdivision vertices removed is an elimination tree for~$G$ of depth at most $k$.
\end{proof}

\begin{corollary}
    \textsc{Elimination Distance to Dominated Clusters} does not admit a polynomial kernel with respect to parameter $k$ even for fixed parameters $d=0$ and $c=2$. 
\end{corollary}

\section{Conclusion}

We have studied the \DCD problem and resolved the open question by Bentert et al.~\cite{bentert2024breaking} whether the problem is fixed-parameter tractable with respect to the parameters \mbox{$k+d+c$}, by proving uniform fixed-parameter tractability even for the smaller parameter $k+d+\ell$, where $\ell$ is the semi-ladder index of the input graph.
We also introduced the more general \EDDC problem and almost fully classified its parameterized and kernelization complexity. 
The most interesting missing part of the classification is whether the case $d=0$, which corresponds to \textsc{Treedepth}, is \textsf{NP}-complete on graphs of bounded maximum degree. We conjecture that this is the case.
Our proofs combine the main tools for addressing cut problems, the decomposition theorem into unbreakable parts by Cygan et al.~\cite{cygan2020randomized} with the main tools for addressing domination-type problems~\cite{FabianskiPST19arxiv}.

\bibliography{../ref}

\appendix
\newpage

\section{Algorithm on \baggraphs}\label{sec:proof-dcd-bags}
This section is devoted to the proof of \cref{prop:dcd-bag}.

\DcdBag*

It closely follows from the proof of~\cref{thm:algo-simple-unbreak} in \cref{sec:unbreakable-bd-semi-ladder}. The main differences are that we now work on an annotated version of the problem, and that the graph might not be unbreakable, only the set of interior vertices are (see \cref{def:baggraphs}).

\begin{definition}
	\label{def:col-skeleton}
	Let $G = (I,E)$ be a $(q,k,d)$-\baggraph, $F$ a set with $E\subseteq F$, $R$ and $B$ additional colors on interior vertices. We call a set $S$ a \emph{skeleton for annotated dominated cluster deletion} with parameters~$k$ and $d$ if there is a superset $S'\supseteq S$ of vertices satisfying:
	\begin{itemize}
		\item $|S'|\le k$,
		\item $S'\subseteq G-F$,
		\item every connected component of $G-S'$ has a Red-Blue dominating set of size at most $d$, and
		\item $S$ contains exactly the vertices of $S'$ that have at least one neighbor in a component $C$ of $G-S$ where $|C\cap I|\le q$, and $C$ contains at least one Red vertex.
	\end{itemize}
\end{definition}

We now define an operation that will be necessary to obtain a statement analogous to~\cref{lem:skeleton-dom-set}.
\begin{definition}\label{def:close-sat}
	For a \baggraph $G=(E,I)$, we say that two interior vertices $u,v$ are {\em close} if either $u$ and $v$ are adjacent, or if there is a path from $u$ to $v$ that only uses exterior vertices.

	We then define the {\em $q$-saturation $\sat_q(u)$} of an interior vertex $u$ as the set of all interior vertices that are close to $u$ if there are at most $q$ of them; or as the singleton $u$ otherwise.
	For a set $X$, $\sat_q(X):= \bigcup\limits_{u\in X}\sat_q(u)$.
	
	For an integer $i$, we define the {\em $q$-saturation at distance $i$} $\sat_q^i(X)$ of a vertex set $X$ as $\sat_q(X)$ if~$i=1$, and $\sat_q(\sat_q^{i-1}(X))$, i.e.~we apply the $q$-saturation operator $i$ times.
\end{definition}
 
Intuitively from an interior vertex, we take all the interior vertices that are adjacent or connected through exterior vertices, but only if there are few of them (at most $q$).

\begin{lemma}\label{lem:adcd-smal-dom}
	Let $q,k,d$ be integers and $G$ be a $(q,k,d)$-\baggraph that is a positive instance to \ADCD, then $G$ has a $3qd$ Red-Blue dominating set.
\end{lemma}
\begin{proof}
	Let $(I,E)$ be the partition of the $(q,k,d)$-\baggraph $G$, and $S$ be a solution of \ADCD, i.e.~every connected component of $G-S$ has a $d$-Red-Blue dominating set. Let $C_0$ be the only connected component of $G-S$ that contains more than $q$ vertices of $I$. In $G-C_0$, there are at most $q$ interior vertices, which together with the~$qd$ Red exterior vertices amount to at most $2qd$ Red vertices in $G-C_0$. So the Blue dominating set for $G-C_0$ is of size at most $2qd$ as well. Together with the $d$ Blue vertices dominating all Red vertices of $C_0$, we have a $3qd$ Red-Blue dominating set.
\end{proof}

\begin{lemma}[analogous to \cref{lem:skeleton-dom-set}]
	Let $G=(I,E)$ be a $(q,k,d)$-\baggraph with colors $F,R,B$, let $S$ be a skeleton for \ADCD with parameters $k,d$, and let $X$ be a~$3qd$ Red-Blue dominating set for $G$. Then there is a vertex of $v\in S$ that is in $sat_q^q(X)$.
\end{lemma}
\begin{proof}
	If there is a vertex of $S$ in $X$, we are done as $X\subseteq\sat_q^q(X)$.
	Otherwise, let $S'\supseteq S$ witnessing that $S$ is a skeleton, let $v$ be a vertex of $S$ and $C$ the connected component of $G-S'$ witnessing that $v\in S$, i.e.~$|C\cap I|\le q$, $C$ contains a Red vertex, and $v$ has a neighbor is $C$.

	Let $u$ be a Blue vertex of $X$ that dominate a Red vertex of $C$. If $u$ is in $S$ then we are done. Otherwise, $u$ is in $C$ and there is a path in $C$ from $u$ to $v$. Each interior vertex $w$ of this path is in $C$ and every vertex close to $w$ is in $C$ or $S$, so there are at most $q$ close vertices and all are in $\sat_q(w)$. This concludes that $v\in\sat^q_q(u)$, hence $v\in \sat^q_q(X)$.
\end{proof}

\begin{definition}[Skeleton algorithm for \baggraph]\label{def:colored-skeleton}
	The {\em annotated skeleton-algorithm} with parameters $q,k,d$ over a $(q,k,d)$-\baggraph $G$ returns a family of sets of vertices, each set of size at most $k$.
	
	First, it computes a dominating set $X$ of size at most $3qd$ (returning the empty family if there is no such set).
	Second, it computes the set $\sat^q_q(X)$.
	Third, it calls the algorithm with parameters $q,k-1,d$ on $G$ (and get a family $\Ff_0$), and on $G-v$ for every $v$ in $\sat^q_q(X)$ (and get a family $\Ff_v$).
	And finally, for every $v$ in $\sat^q_q(X)$ add $v$ to every set in $\Ff_v$, and return the union of all the families.
\end{definition}

\begin{remark}
	Note that for any interior vertex $v$, $G-v$ is a $(q,k-1,d)$-\baggraph. Furthermore, if the input graph is a positive instance of \ADCD, and if the algorithm guesses correctly a vertex $v$ of a solution, then $G-v$ is also a positive instance to \ADCD with parameter $(k-1,d)$. So the algorithm can repeat until it guesses a skeleton $S$ of size at most $k'$ in time \mbox{$f(q,k,d,t(G))\cdot |G|^{\Oof(1)}$}, where $t(\cdot)$ is a parameter satisfying~\cref{para-dom-set}.
\end{remark}

It remains to verify that $S$ can indeed be enlarged to a set $S'$ of size at most $k$ with the property that every connected component of $G-S'$ has a $d$-Red-Blue dominating set.

Note that by definition, there is at most one component $C_0$ of $G-S$ that contains more than $q$ interior vertices. So in $G-C_0$, there are at most $3qd+q$ many Red vertices ($3qd$ for the exterior vertices and $q$ for the interior ones).
What remains to be done is hence the following. 

\begin{enumerate}
	\item Find an optimal solution for $G-S-C_0$.
	This can be done in time $f(k,d)$, because it contains at most $q$ interior vertices and only these can be deleted (so we can try all possibilities), and it contains at most $3qd+q$ Red vertices. As we can treat indifferently two Blue vertices with the same Red neighborhood, and since there are at most $2^{O(qd)}$ possible Red neighborhoods in time $f(q,k,d)$ we can go through all possibilities to dominate the Red vertices.

	Each such possibility yields a value $k'$ of vertices that need to be deleted.
	If $k'>k-s$ where $s=|S|$, we conclude that this $S$ is not a skeleton because there is no possible $S'$.

	\item Test whether an additional set $W\subseteq C_0-F$ of $k''=k-s-k'$ vertices that can be deleted from $C_0$ such that $G[C_0-W]$ has a $d$-Red-Blue dominating set.
	Hence, the remaining problem is an instance of \textsc{Partial Red-Blue Domination} with parameter $k''$ and $d$.
	If $G$ is a positive instance of \ADCD, witnessed by some set $S'$, and if $S$ is a skeleton of $S'$, such a set $W$ exists (as $S'\cap C_0$ is a candidate).
	If $t(\cdot)$ is a parameter satisfying~\cref{para-dom-set}, we can then test for the existence of such $W$ in time $f(q,k,d,t(G))\cdot|G|^{\Oof(1)}$. If no solution is found, then this $S$ is not a skeleton.
\end{enumerate}

This concludes the proof of \cref{prop:dcd-bag} and therefore \cref{sec:proof-dcd-bags}.

\pagebreak
\section{Algorithm through unbreakable tree decomposition}\label{sec:dcd-tree-decomposition}

In this section, we assume that we have a fixed graph $G$, and a $(q,k)$-unbreakable tree decomposition~$(T,\bag)$ of $G$. See~\cref{sec:prelims} for the definition of an unbreakable tree decomposition, and \cref{thm:CyganLPPS19} (\cite[Theorem 10]{CyganLPPS19}) for the proof of its existence.

The goal of this section is to prove \cref{main-theorem-DCD} using the fact that \ADCD is FPT on \baggraphs (\cref{prop:dcd-bag}).

\maintheoremDCD*

\subsection{Marks and unbreakable \baggraph}\label{seq:mark-baggraphs}

\begin{definition}\label{def:mark2}
	Given a graph $G$ (with colors $F,R,B$), a subset of vertices $A$, and integers $k,d$, we define a $(k,d)$-{\em mark} of $A$ in $G$ as a tuple of the form:
	\begin{itemize}
		\item disjoint subsets $S_A,D_A,U_A$ of $A$,
		\item an integer $k_A\le k$,
		\item a partition $\Pp$ of $A-S_A$, and
		\item for each element $P$ of the partition, an integer $p\le d$.
	\end{itemize}\medskip
	A mark is said to be {\em realized} if there is a set $S\subseteq V(G)-F$ of at most $k$ vertices such that:
	\begin{itemize}
		\item $S_A = S\cap A$,
		\item $k_A = |S|-|S_A|$,
		\item for every connected component $C$ of $G-S$ either :
		\begin{itemize}
			\item $C\cap A = \emptyset$ and $C$ has a Red-Blue dominating set of size $d$, or
			\item $C\cap A \neq \emptyset$, there is a $P\in\Pp$ such that $P=C\cap A$. Furthermore, there must exist a subset $D_C\subseteq C\cap B$ of size $p$ such that
			$D_C\cap P = D_A\cap P$, and
			$D_C$ dominates $R\cap(C-U_A)$.
		\end{itemize}
	\end{itemize}
\end{definition}
Intuitively, we mark how the set $A$ overlaps with the deleted vertices and the Blue dominating sets. The Blue dominating sets must dominate all Red vertices of $G-A$, and if some Red vertices of $A$ are not dominated, it is written so on $U_A$ (for undominated).

\begin{definition}\label{def:profile}
	Given a graph $G$, a subset of vertices $A$, and integers $k,d$, the $(k,d)$-{\em profile} of $A$ in~$G$ is the list of all realized $(k,d)$-marks of $A$ in $G$.
\end{definition}

\begin{lemma}\label{lem:size-profile}
	The $(k,d)$-profile of a set $A$ of size $|A| = 2^{\mathcal{O}(k)}$ in a graph $G$ has size at most~$d^{2^{\O(k)}}$.
\end{lemma}
\begin{proof}
	There are $\mathcal{O}(4^{|A|})$ possibilities to split the set $A$ into the disjoint subsets $S_A,D_A,U_A$, and the rest, and $k_A$ is an integer bounded by $k$.
	Furthermore, there are at most $|A|$ elements in the partition, each is given an integer bounded by~$d$.
	This results in $\mathcal{O}(d^{|A|})$ possibilities.
	
	Assuming that $|A|=2^{\mathcal{O}(k)}$, this yields profiles of size 
	$d^{2^{\O(k)}}$.
  \end{proof}

The overall algorithm will inductively compute the profile of each cone of the tree decomposition (with respect to its adhesion). The profile of the root bag will contain the information to answer whether the graph is a positive instance to the \textsc{Annotated Dominated Cluster Deletion} problem. The main proof of this section is that we can compute the profile of a cone given the profile of every subtree. To achieve that we need to add some structure to the bags of the tree decomposition.

\begin{definition}\label{def:gadget}
	A {$d$-gadget} is the graph $H$ composed of $2d+1$ vertices $a, b_1,\ldots,b_d,b'_1\ldots,b'_d$, where for every $i\le d$, $b_i$ is adjacent to $a$ and $b'_i$.

	A $d$-gadget is {\em plugged} to a set $A$ by making every vertex of $A$ adjacent to $a$.
\end{definition}

Note that $H$ is the one-subdivided star with $d$ branch. With a slight abuse of notation, we also call $d$-gadget, any graph that are $d'$-gadget with $d'\le d$.

\begin{definition}\label{def:bgraph}
	Let $(T,\bag)$ be a regular $(q,k)$-unbreakable tree decomposition of a graph $G$, $S\subseteq\cone(x)$ a set of at most $k$ vertices, and $d$ an integer.
	We note {\em $\bgraph_S(x)$} of a node $x\in V(T)$ any graph that is obtained from 
	$G[\cone(x)-S]$ by
	\begin{itemize}
		\item contracting each connected component $C$ of the set $\cmp(y)-S$ that has a neighbor in $\adh(y)-S$ into a $d$-gadget $H_y^C$ and plugging $H_y^C$ to the vertices of $\bag(x)-S$ that are neighbors of $C$, and
		\item adding at most $q$ many $d$-gadget plugged to a subset of the adhesion of $x$.
	\end{itemize}
\end{definition}

\begin{lemma}
	Every $\bgraph_S(x)$ is a $(q,k-|S|,d)$-\baggraph.
\end{lemma}
\begin{proof}
	The vertices of $\bgraph_S(x)$ are split into $I$ the vertices of $\bag(x)$, and $E$ the union of all the $d$-gadgets. Clearly, every connected component of $G-I$ is a $d$-gadget and therefore has at most $2d+1$ vertices. Furthermore, this $d$-gadget is plugged to a subset of the adhesion of $x$ or the adhesion of a child of $x$, and therefore has at most $q$ neighbors in $I$.

	For the last point, let $L,R$ be a separation of $\bgraph_S(x)$ of order at most $k-|S|$, then extend $L,R$ into $L',R'$ by first removing the $d$-gadget plugged to the adhesion of $x$, then adding $S$ to both $L$ and $R$, and finally replacing every $d$-gadget plugged to a subset of $\adh(y)$ by the component of $\cone(y)$ it represents.

	We obtain that $(L',R')$ is a separation of order $k$ of $\cone(x)$, and therefore $|L'\cap \bag(x) |\le q$ or $|R'\cap \bag(x) |\le q$, hence with $I=\bag(x)$, either $|L\cap I|\le q$, or $|R\cap I|\le q$.
\end{proof}

Note that $\bgraph_S(x)$ is not just one graph, but rather a family of graphs. Nonetheless, it is convenient to write $|\bgraph_S(x)|$ and $\|\bgraph_S(x)\|$ to denote the maximum amount of vertices (resp. of edges) that one of the $\bgraph_S(x)$ can have. We now provide bounds for these numbers.

\begin{lemma}\label{lem:size-bgraph}
	Let $(T,\bag)$ be a regular $(q,k)$-unbreakable tree decomposition of a graph $G$. Then 
	$\sum\limits_{x\in V(T)}|\bgraph(x)|\le (4d+5)q|G|$ and
	$\sum\limits_{x\in V(T)}\|\bgraph(x)\|\le \|G\|+((4d+5)q)^2|G|$.
\end{lemma}

\begin{proof}
	For every $x\in V(T)$, $\bag(x)$ is the disjoint union of $\adh(x)$ and $\mrg(x)$.
	Every vertex of~$G$ appears in exactly one margin of one node in the tree decomposition and each node of the tree decomposition has an adhesion of size at most $q$.
	Furthermore, each node is the child of at most one node and each connected component of a child can be divided in at most $q$ many connected components. Additionally, $d$-gadgets have size at most $2d+1$, and $|T|\leq |G|$, hence:
	\[\sum\limits_{x\in V(T)}|\bgraph(x)|\leq (2d+2)q|T|+|G|+(2d+2)q|T|\leq (4d+5)q|G|\]
  
	For the second bound, we consider the total number of vertices and edges in all bag graphs.
	Each edge in the margin of a node in the tree decomposition corresponds to exactly one edge in the graph~$G$.
	Additionally, there are at most $q\cdot|\bgraph(x)|$ many edges in each bag graph that are incident to a vertex in the adhesion and there are at most $(2qd)\cdot|\children(x)|$ many edges in the $d$-gadgets plugged to the adhesion of the children of $x$ and $(2qd)$ edges in the $d$-gadgets plugged to the adhesion of $x$.
	\begin{align*}
		\sum\limits_{x\in V(T)}\|\bgraph(x)\|
		&\leq\sum\limits_{x\in V(T)}|\bgraph(x)|+\|G\|+2dq\cdot\sum\limits_{x\in V(T)}(|\bgraph(x)|+|\children(x)|)\\
		&\leq (4d+5)q|G|+\|G\|+2dq((4d+5)q|G|+|T|)\\
		&\leq \|G\|+((4d+5)q)^2|G|
	\end{align*}
	This concludes the proof.
  \end{proof}

In the next lemma, we show that the ladder index of these bag graphs is bounded (assuming that it was already bounded in the original graph). This together with the fact that the semi-ladder index has \cref{para-dom-set} implies that the semi-ladder index also has~\cref{para-bound-bgraph}.

\begin{lemma}\label{lem:para-bgraph}
	Let $\ell$ be an integer and $(T,\bag)$ be regular $(q,k)$-unbreakable tree decomposition of a graph $G$ with semi-ladder index at most~$\ell$.
	Then, for every node $x\in V(T)$  every bag graph $\bgraph_S(x)$ has semi-ladder index at 
	most~$q+\ell+4$.
\end{lemma}
\begin{proof}
	Let $a_1,\ldots,a_t,b_1,\ldots,b_t$ be a semi-ladder of $\bgraph_S(x)$. 
	Note that in a $d$-gadget, only one vertex has degree greater than 2. Hence, the other vertices can only be among $a_1,a_2,a_3$ and $b_t,b_{t-1},b_{t-2}$ and since the only high degree vertex of a $d$-gadget has at most $q$ neighbors that are in $\bag(x)$, we conclude that the vertices from the $d$-gadget can only be among $a_1,\ldots,a_{q+4}$, and symmetrically in~$b_{t-q-3},\ldots,b_t$.
	Therefore, $a_{q+5},\ldots a_t,b_1,\ldots,b_{t-q-4}$ is a semi-ladder of $G$. 
	So~$t-(q+5)+1\le \ell$, hence $t\le q+\ell+4$.
\end{proof}

We state the next proposition with arbitrary parameter $t(\cdot)$ satisfying \cref{para-bound-bgraph}. By the previous lemma, the semi-ladder index $\ell$ can replace $t(\cdot)$.

\begin{proposition}\label{main-dp-algo-ADCD}
Given a $(q,k)$-unbreakable tree decomposition of a graph $G$, a node $x$ of the tree decomposition and, for each child $y$ of $x$, the $(k,d)$-profile of $\adh(y)$ in $\cone(y)$, we can compute the $(k,d)$-profile of $\adh(x)$ in $\cone(x)$,
in time $f(q,k,d,t(G))\cdot|\bgraph(x)|^{\Oof(1)}$, where~$t(\cdot)$ is a parameter satisfying \cref{para-bound-bgraph}, or alternatively in time $f(q,k,d)\cdot|\bgraph(x)|^d$.
\end{proposition}

Note that by~\cref{lem:size-bgraph}, the sum of the size of all bag graphs is linear in the size of the graph so that~\cref{main-dp-algo-ADCD} yields~\cref{thm:CDCD}, and therefore~\cref{main-theorem-DCD}.

\subsection[Algorithm for Proposition 49]{Algorithm for~\cref{main-dp-algo-ADCD}}\label{subsec:algo-ADCD}

Let $G$ be a graph, $(T,\bag)$ a $(q,k)$-unbreakable tree decomposition and $x\in V(T)$ a node of this decomposition. We assume that we have access to the profile of $\adh(y)$ in $\cone(y)$ for every child~$y$ of $x$. We fix a $(k,d)$-mark $M_x$ of $\adh(x)$ in $\cone(x)$ and test whether it is realized.
In order to simplify a bit the notation, given a child $y$ of $x$ we will simply call a mark of $y$ a $(k,d)$-mark of $\adh(y)$ in $\cone(y)$.

\medskip\noindent
{\bf Initialization.}
We now design a backtracking algorithm that start with a partial solution\linebreak \mbox{$W = \{S=\emptyset,s=0,D=\emptyset,$} $c=0\}$. Intuitively, $S$ (resp.\ $D$) contains the vertices of $\bag(x)$ to be deleted (resp.\ picked as dominators), and $s$ counts the vertices to be deleted in the descendant of~$x$.
Finally, the role of $c$ is to count the dominators from the children of $x$ that are in a connected component overlapping both $x$ and $y$. 

Before going to the children, let $S_x,D_x,U_x,k_x,\Pp_x, (p_i)_{P_i\in \Pp_x}$ be the mark selected for $x$. We add~$S_x$ to $S$, $k_x$ to $s$ and $D_x$ to $D$. 

%
\medskip\noindent
{\bf Branching through children's marks.}
We now go over all children $y$ of $x$ and update our partial solution $W$. There are two ways our partial solution is updated, depending on the marks present in the profile of $y$. If the following mark is realized

$$S_y = \adh(y)\cap S ~~;~~ 
D_y = \adh(y) \cap D ~~;~~
k_y=0 ~~;~~
\bigwedge\limits_{P_i\in\Pp_y}p_i=0,$$
then we do not update our partial solution, this child is already taken care of, and we move to the next child of $x$.

Otherwise, our algorithm branches over all the realized marks of $y$. For each branch, considering a mark $M'$,  we check the consistency of $M'$ with our partial solution, and if it is consistent, we update our connected component labeling and our partial solution.

\pagebreak\noindent The checks are:
\begin{itemize}
	\item $S\cap \adh(y)\subseteq S_y \subseteq G-F$, i.e.~agrees on the vertices being already deleted and none are in $F$.
	\item $D\cap \adh(y)\subseteq D_y\subseteq B$, i.e.~agrees on the vertices already picked to be dominators, and they are Blue vertices.
	\item $D\cap S_y = S \cap D_y = \emptyset$, i.e.~a vertex cannot be both deleted and a dominator.
\end{itemize}

For the updates, first, add $A_S$ to $S$, and $k_y$ to $s$. Second, for every $P_i\in{\Pp}_y$ add $p_i$ to $c$.

If any of the previous checks failed, as well as if either $|S|+s > k$, or if $|D|+c>qd$, we terminate this branch of the algorithm.

\medskip\noindent
{\bf Reducing to bag graphs.}
We have $S,s,D,c$ fixed and one selected mark for every child $y$ and one for $x$.
We compute a \baggraph $H$ as follows:

First, starts with $H:=G[\bag(x)-S]$, taking the edges and colors $F,R,B$ from $G$.

Second, for every child $y$ let $S_y,D_y,U_y,k_y,\Pp_y$ and an integer $p$ for each $P\in\Pp_y$ be the selected mark. Then:
\begin{itemize}
	\item remove $\adh(y)-U_y$ from $R$ {\it(these vertices are dominated in $y$)},
	\item remove $N(D_y)$ from $R$ {\it(we mark these vertices as dominated)}, and
	\item for every $P_i\in\Pp_y$ plug the $p_i$-gadget to the vertices of $P_i$.
\end{itemize} 

Finally, let  $S_x,D_x,U_k,k_x,\Pp_x,$ and a $p$ for each $P\in\Pp_x$ be the selected mark for $\adh(x)$.
\begin{itemize}
	\item Remove $N(U_x)$ from $B$ {\it($U_x$ shall not be dominated)},
	\item remove $N(D_x)$ from $R$ {\it(we mark these vertices as dominated)}, and
	\item for every $P\in\Pp_x$ plug the $(d-p)$-gadget to the vertices of $P$.
\end{itemize} 

 We are now done building a \baggraph $H$, of the form $\bgraph_S(x)$. We now conclude by testing whether $H$ is a positive instance to $(k_x-|S|-s,d)$-\ADCD with \cref{prop:dcd-bag}.

\begin{claim}[Correctness of the algorithm]
	The mark $M_x$ of $\adh(x)$ in $\cone(x)$ is realized if and only if $H$ is a positive instance to $(k_x-|S|-s,d)$-\ADCD. 
\end{claim}
\begin{proof}
	For the first direction, assume that $H$ is a positive instance to $(k_x-|S|-s,d)$-\ADCD, then the algorithm of \cref{prop:dcd-bag} produced a set $S'$ of size $k_x-|S|-s$, which together with $S$ and the sets $S_y$ (of cumulative size $s$) selected for the children satisfies that every connected component of $\cone(x)-S-S'-\bigcup\limits_{y\in\children(x)}S_y$ that does not intersect $\adh(x)$ has a Red-Blue dominating set of size $d$. For the component intersecting $\adh(x)$ the Red-Blue dominating set of size $d$ must contain the $d-p$ Blue vertices of the $(d-p)$-gadget; hence the component has a Red-Blue dominating set of size $p$.

	For the second direction, assume that the mark is realized, hence there is a set $\hat S$ on $\cone(x)$ witnessing it this mark is realized, together with a set of Blue vertices $D_C$ for every connected component of $\cone(x)-\hat S$. When branching over the marks of the children of $x$, the algorithm select the mark for which $\hat S\cap \cone(y)$ is a witness. The built graph $H$ is then a positive instance witnessed by $\hat S \cap \bag(x)$.
\end{proof}

\pagebreak
\subsection{Running time and Black-White trees}\label{subsec:BW}

In this section, we explore the tree produced by our backtracking algorithm. We will show that this tree must have a specific structure that we call Black-White (with additional parameters), and such trees must be of linear size and have a constant number of leaves.

\begin{definition}
	A tree is called {Black-White} if every node of the tree is colored either Black or White, and every non-leaf node either has exactly one White child or only Black children.

	The {\em degree} of a \BW is the maximum number of Black children a node has in the tree. The {\em budget} of a \BW is the maximum number of Black vertices in a root-to-leaf path. 
\end{definition}

\begin{lemma}\label{lem:rb-size}
	Any \BW of degree $\al$, budget $\be$, and depth $n$ has size at most $\al^\be\cdot n$ and the number of leaves is at most $\al^\be$.
\end{lemma}
\begin{proof}
	The bound on the number of leaves is straightforward.
	For the size, notice that whenever a White vertex has some Black children, it only increases the size of the tree to switch them (i.e.~the parent of the White vertex now has several Black children, each with one White child), and this operation maintains the degree, budget, and depth of the tree. Therefore, the maximal Black-White tree is a complete Black tree of degree $\al$ and of depth $\beta$, where each leaf is attached to a White path of length $n-\beta$, for which the size is at most $\al^\be\cdot n$.
\end{proof}

When doing the backtracking, it is key that if a child is satisfied with the given partial solution, we do not try to branch and make an impactful decision. Otherwise, the tree of partial solutions grows too much and is not linear. In the context of \BW, that would be like allowing one White child and some Black children.
\begin{lemma}
	The branching tree of the backtracking algorithm of \cref{subsec:algo-ADCD} can be labelled to be a \BW of degree $d^{2^{\Oof(k)}}$ and of budget $(k+qd)$.
\end{lemma}
\begin{proof}
	When we branch over the children we only have $d^{2^\Oof(k)}$ many marks to consider (see \cref{lem:size-profile}). This bounds the degree of our \BW.

	Furthermore, every time that we branch, this increases at least one of $|S|$, $s$, $|D|$, or $c$. As we terminate the algorithm as soon as $|S|+s>k$ or $|D|+c>qd$, the budget remains bounded by~$k+qd$.
\end{proof}

Going back to the algorithm for \cref{main-dp-algo-ADCD} presented in \cref{subsec:algo-ADCD}, we just showed that the number of possible branches to consider is only a function of $q,k$, and $d$. And since each branch then perform a single call to \cref{prop:dcd-bag}, this concludes the proof of \cref{thm:CDCD}. 

\section{Annotated version for \EDDC}
\label{sec:aeddc}

Similarly to \cref{sec:annotated}, we extend \cref{main-theorem-EDDC} to an annotated version.
\begin{definition}[\AEDDC]
	Given a graph $G$, integers $k$, $d$, and vertex sets $F,R,B$, the $(k,d)$-\AEDDC problem is to find a tree-structured set $S\subseteq G-F$ of elimination depth $k$ such that every connected component $C$ of $G-S$ has a Red-Blue dominating set of size at most $d$ (i.e.~$d$ Blue vertices dominating all Red vertices).
\end{definition}

\begin{definition}
	Given a graph $G$, a subset of vertices $A$, and integers $k,d$, we define a $(k,d)$-{\em extended-mark} 
	of $A$ in $G$ as a tuple of the form:
	\begin{itemize}
		\item disjoint subsets $L_A^1,\ldots,L_A^k,D_A,U_A$ of $A$,
		\item a partition $\cal P$ of $A-L_A$, where $L_A = L_A^1 \cup \ldots \cup L_A^k$, and
		\item for each element $P$ of the partition:
		\begin{itemize}
			\item an integer $p\le d$,
			\item booleans $r^1,\ldots,r^k$.
		\end{itemize} 
	\end{itemize}

	An extended-mark is said to be {\em realized} if there is a set $S$, partitioned into sets $L^1,\ldots,L^k$ of vertices such that:
	\begin{itemize}
		\item for every $i\le k$ and vertices $u,v$ in $L^i$ every $u-v$ path must use a vertex in $L^{j<i}$,
		\item $L_A^i = L^i\cap A$, and
		\item for every connected component $C$ of $G-S$ either:
		\begin{itemize}
			\item $C\cap A = \emptyset$ and $C$ has a Red-Blue dominating set of size $d$, or
			\item $C\cap A \neq \emptyset$ and there is a $P\in\cal P$ such that $P=C\cap A$. Furthermore, there must exist a subset $D_C\subseteq C\cap B$ of size $p$ such that 
			$D_C\cap P = D_A\cap P$ and $D_C$ dominates $R\cap(C-U_A)$.
			
			And $r^i=True$ if and only if there is a path from any vertex of $P$ to any vertex of $L^i$ that is not using vertices of $L^{j<i}$.
		\end{itemize}
	\end{itemize}
\end{definition}

The first point corresponds to the fact that $S$ is tree-structured with elimination depth $k$. The sets $L^i_A$ correspond to the layer of the structure.

\begin{definition}
	Given a graph $G$, a subset of vertices $A$, and integers $k,d$, the $(k,d)$-{\em extended-profile} of $A$ in $G$ is the list of all realized $(k,d)$-extended-marks of $A$ in $G$.
\end{definition}

The proof of the next observation is similar to the one of~\cref{lem:size-profile}.
\begin{observation}
	The $(k,d)$-extended-profile of a set $A$ of size at most $q$ in a graph $G$ has size at most $(d+k)^{2^{\O(k)}}$. 
\end{observation}

We are now going to work with the same definition of gadget and bag graphs (see \cref{def:gadget,def:bgraph}).

\begin{proposition}\label{main-dp-algo-EDDC}
	Given a $(q,k)$-unbreakable tree decomposition of a graph $G$, a node $x$ of the decomposition and, for each child $y$ of $x$, the $(k,d)$-extended-profile of $\adh(y)$ in $\cone(y)$, we can compute the $(k,d)$-extended-profile of $\adh(x)$ in $\cone(x)$,
	in time either $f(q,k,d)\cdot|\bgraph(x)|^d$, or $f(q,k,d,t(G))\cdot|\bgraph(x)|^{\Oof(1)}$ where~$t(\cdot)$ is a graph parameter satisfying \cref{para-bound-bgraph}.
\end{proposition}

Equivalently, the running time is $f(q,k,d,t_1(\bgraph(x)))\cdot|\bgraph(x)|^{\Oof(1)}$ where~$t_1(\cdot)$ is a graph parameter satisfying \cref{para-dom-set}.
Note that by~\cref{lem:size-bgraph}, the sum of the size of all bag graphs is linear in the size of the graph so~\cref{main-dp-algo-EDDC} yields our main result,~\cref{main-theorem-EDDC}.

\subsection[Algorithm for Proposition 58]{Algorithm for \cref{main-dp-algo-EDDC}}

Let $G$ be a graph, $(T,\bag)$ a $(q,k)$-unbreakable tree decomposition and $x\in V(T)$ a node of this decomposition. We assume that we have access to the extended-profile of $\adh(y)$ in $\cone(y)$ for every child~$y$ of $x$. We fix a $(k,d)$-extended-mark $M_x$ of $\adh(x)$ in $\cone(x)$ and test whether it is realized.
In order to simplify a bit the notation, given a child $y$ of $x$ we will simply call an {\em emark of~$y$} a $(k,d)$-extended-mark of $\adh(y)$ in $\cone(y)$.
  
\paragraph*{Initialization}
We now design a backtracking algorithm that start with a partial solution \mbox{$W = \{S^1=\ldots=S^k=\emptyset,$} $D=\emptyset, s^1=\ldots=s^k=0, c=0\}$. Intuitively, $S^i$ (resp.\ $D$) contains the vertices of $\bag(x)$ in the $i$th layer of the tree structure (resp.\ picked as dominators), and $s^i$ counts the vertices to be deleted in the descendant of $x$ that are in layer $L^i$ and reachable by a path not using $L^{j<i}$.
Finally, the role of~$c$ is to count the dominators from the children of $x$ that are in a connected component overlapping both $x$ and $y$.

Before going to the children, let $L^1_x,\ldots,L^k_x,D_x,U_x,\Pp_x, (p_i)_{P_i\in \Pp_x},r^1_x,\ldots,r^k_x$ be the mark selected for $x$. We add~$L^i_x$ to $S$, and $D_x$ to $D$.

%

\paragraph*{Children}
We now go over all children $y$ of $x$ and update our partial solution $W$. There are two ways our partial solution is updated, depending on the extended-marks present in the extended-profile of $y$. If the following extended-mark is realized

$${L^i_y} = \adh(y)\cap S^i ~~(\text{for all } i\le k);~~ 
D_y = \adh(y) \cap D ~~;~~$$
$$r^i=False ~~(\text{for all } i\le k);~~ 
\bigwedge\limits_{P_j\in\Pp_y}p_j=0,$$
then we do not update our partial solution, this child is already taken care of, and we move to the next child of $x$.

Otherwise, our algorithm branches over all the realized extended-marks of $y$. For each branch, considering an extended-mark $M_y$, we check the consistency of $M_y$ with our partial solution, and if it is consistent we update our connected component labeling and our partial solution.
The checks are :
\begin{itemize}
	\item $S^i\cap \adh(y)\subseteq {L^i_y}$, i.e.~agrees on the vertices being already deleted, and their layer.
	\item $D\cap \adh(y)\subseteq D_y$, i.e.~agrees on the vertices already picked to be dominators.
	\item $S^i \cap D_y = D\cap {L^i_y} = S^i\cap {L_y^{j\neq i}} = \emptyset$, i.e.~a vertex cannot be both deleted and a dominator and if deleted only on a single layer.
\end{itemize}

As for the updates, first, add $L_y^i$ to $S^i$, and increment $s^i$ if $r^i = True$. Second, for every $P_i\in{\Pp_y}$ add $p_i$ to $c$.

If any of the previous checks failed, as well as if either $|S^i|+s^i > qk$ (for any $i\le k$), or if~$|D|+c>qd$, we terminate this branch of the algorithm.

%
\paragraph*{Bag graphs and skeletons}
We now have a fixed set $S=\bigcup\limits_{i\le k}S^i$, and similarly to \cref{subsec:algo-ADCD}, we compute the bag graph $H_x=\bgraph_S(x)$.
Similarly to \cref{sec:proof-dcd-bags} (see \cref{lem:adcd-smal-dom}) positive instances have a small Red-Blue dominating set. This transfers to the elimination distance variation, as it was the case in \cref{lem:neg-instance-eddc} for unbreakable graphs.
So the colored version of skeleton-algorithm (see~\cref{def:colored-skeleton}) can also be used here.

Then we therefore compute every skeleton $S'$ with parameter $k$ and $3qd$ of $H_x$. We then guess in which $S^i$ we assign each element of $S$ (which modifies $S$).
We terminate this branch of the algorithm if $|S^i |+s_i>qk$, or if $S'$ contains a vertex that lies in the adhesion of a child of~$x$ (meaning that the selected mark for this child is inconsistent with the selected skeleton).

This step can be performed in time $f(q,k,d,t_1(H_x))\cdot |H_x|$, where $t_1(\cdot)$ satisfies \cref{para-dom-set}, meaning in time $f(q,k,d,t_2(G))\cdot |H_x|$, where $t_2(\cdot)$ satisfies \cref{para-bound-bgraph}. 
Alternatively, in time $f(q,k,d)\cdot |H_x|^{\O(d)}$, in which case we also store the set $D'$ that dominates the large connected component. See~\cref{lem:poly-comput-skeleton}, which can be adapted to bag graphs.

\paragraph*{Fixing the small connected components}
We now define $C_0$ as the only connected component of $H_x-S$ that contains more than $q$ vertices of $\bag(x)$.
We branch over every choice $S''\subseteq \bag(x)-(S'\cup C_0)$ such that $S''\cup S$ still has a tree structure of elimination depth $k$.
This creates connected components $C_1,\ldots,C_{q}$ of $\bag(x)-(S\cup S'' \cup C_0)$.
We guess for each vertex of $S''$ in which $S_i$ it is assigned.

For every $i>0$, we define $D_i:=D\cap C_i \cup D'_i$, where $D$ is the set computed at the end of the {\it Children} step and $D'_i$ is a guessed subset of at most $d$ Blue vertices of $C_i$ (we branch over all possibilities).

\paragraph*{Partial domination in the large connected component}
We now compute a partial dominating set in the large connected component. This is achieved as for \Cref{cor:partial-dom-fpt} for which the colors $F,R,B$ are defined as:
\begin{itemize}
	\item $F$: Starts with $F$ from $G$, add every $d$-gadgets, and the adhesion of every $y$, children of $x$, together with the adhesion of $x$.
	\item $R$: Remove from $R$ every vertex already dominated by the current value of $D$.
	\item $B$: Remove from $B$ the vertices of $\adh(x)$ that are in $U_x$ and their neighborhood (as these vertices shall not be dominated).
\end{itemize}
We check the existence of a solution for every pair of parameters $k'\le k$, and $d'\le d$ (we branch over the choice of $k'$ and $d'$).

Alternatively, we add $D'$ to $D_0$ and check that it dominates $C_0$. In the case $D'$ was computed in time $f(q,k,d)\cdot |H_x|^{\O(d)}$.

\paragraph*{Sanity checks}

We no longer branch and only check whether this branch satisfies our requirements.

First check that in every small connected component $C_{i>0}$, every Red vertex is dominated by~$D_i$. We then check that the connected components of $\cone(x)$ all have small Red-Blue domination sets, i.e.~there could be connected components $C_i$, $C_j$ of $\bag(x)$ that both have a neighbor in a $d$-gadget in $H_x$. We call two such connected components related.

We can then define the {\em related graph} on the connected components by going through all the $d$-gadget. This yields a partition of the connected components where each family is a maximal subset of related connected components.

Then for every $I\subset \{0,\ldots q\}$ of this partition we add up the size of $|D_{i\in I}|$, as well as the relevant~$p_j$ from the selected marks of the children of $x$ that are adjacent to $C_{i\in I}$.

After that, we check the consistency of the $S^1,\ldots,S^k$, i.e.~no two vertices of $S^1$ are connected in the graphs, then no two vertices of $S^2$ can be connected not using $S^1$ etc.
This can be done in linear time: Simply brute force the paths in the small components, and, in the big component, every two vertices are connected not using $S$ (otherwise the large component is split).

We finally check that the current solution matches the selected extended-mark $M_x$ that we are trying to realize. So
we check that the number of deleted vertices in each of the $k$ layer correspond to the values in  $r^i_x$ (for each $i\le k$), and similarly for the number of picked dominators per family of connected components $C_{i\in I}$ for each relevant $I$, and check that the partition $\Pp_x$ agrees on this split.

If all the checks were successful, the algorithm claims that the mark is realized.

\subsection{Correctness and running time}
Both are straightforward adaptation of what was done for \ADCD. For example, \cref{subsec:BW} also exactly explains the running time of the current algorithm.

For correctness, pick a realized $(k,d)$-extended-mark $M_x$ of $\adh(x)$ in $\cone(x)$. Let $\hat S$ be the witness set (together with its tree structure). And for every connected component $C$ of $\cone(x)-S$ a set $D$ whose existence is guarantied as the mark is realized.

When going through the 'Children' steps, there is a branch of the algorithm that always took the extended-mark $M_y$ of $y$ that is realized and for which $S$ and the witness. Similarly, there is a choice of a skeleton that is exactly the skeleton of $S$. The other vertices of $S'$ are either guessed from the small connected component, or with the partial domination in $C_0$.
So the algorithm agrees.

The other direction is straightforward as the algorithm produced the witness sets and checks the existence of its tree structure.

\end{document}